\documentclass[10pt]{iopart}
\usepackage{iopams}  
\usepackage{bm}
\usepackage[]{graphicx}
\usepackage[tight]{units}
\usepackage{color}
\usepackage{ulem}
\usepackage{marvosym}
\usepackage[colorlinks=true,citecolor=blue,linkcolor= blue,linkcolor=blue,urlcolor=blue]{hyperref}
\usepackage{lipsum}
\usepackage{braket}

\begin{document}
\newcommand{\NFH}[1]{\textcolor{red}{(#1)}}
\newcommand{\FLOR}[1]{\textcolor{green}{(#1)}}

\newcommand{\text}[1]{\mathrm{#1}}
\newcommand{\F}[1]{Fig.~\ref{#1}}
\newcommand{\f}[1]{fig.~\ref{#1}}
\newcommand{\Figs}[1]{Figs.~\ref{#1}}
\newcommand{\figs}[1]{figs.~\ref{#1}}
\newcommand{\Eq}[1]{Eq.~\ref{#1}}
\newcommand{\eq}[1]{eq.~\ref{#1}}

\newcommand{\kdep}{$\mathbf{k}$-dependent~}
\newcommand{\qdep}{$\mathbf{q}$-dependent~}

\newcommand{\ESF}{\textsc{\'Eliashberg} spectral function~}
\newcommand{\Fermi}{\textsc{Fermi}~}
\newcommand{\Boltzmann}{\textsc{Boltzmann}~}
\newcommand{\Brillioun}{\textsc{Brillioun}~}

\newcommand{\ko}{\mathbf{k}n}
\newcommand{\kp}{\mathbf{k}^{\prime}m}
\newcommand{\q}{\mathbf{q}\nu}
\newcommand{\kos}{\mathbf{k}_n}
\newcommand{\kps}{\mathbf{k}^{\prime}_m}
\newcommand{\qs}{\mathbf{q}_\nu}
\newcommand{\kk}{\mathbf{k}}
\newcommand{\qq}{\mathbf{q}}
\newcommand{\aaF}{\alpha^{2}F_\mathbf{k}(\omega,\epsilon_\mathbf{k})}
\newcommand{\aaFs}{\alpha^{2}F_\mathbf{k}}

\newcommand{\E}{\epsilon_{\ko}}
\newcommand{\Ep}{\epsilon_{\kp}}
\newcommand{\wq}{\hbar\omega_{\q}}
\newcommand{\Erenorm}{\bar{\epsilon}_{\ko}}
\newcommand{\Ef}{E_\text{F}}

\newcommand{\lk}{\lambda_{\ko}}
\newcommand{\lks}{\lambda_\mathbf{k}}
\newcommand{\lktrs}{\lambda_\mathbf{k}^\text{tr}}
\newcommand{\tautr}{\tau^{\text{tr}}}

\newcommand{\me}{\mathfrak{g}_{\kp,\ko}^{\q}}
\newcommand{\mes}{\left|\me\right|^2}

\newcommand{\delQE}{\delta\left(\E-\Ep\right)}
\newcommand{\delp}{\delta\left(\E-\Ep+\hbar\omega\right)}
\newcommand{\delm}{\delta\left(\E-\Ep-\hbar\omega\right)}
\newcommand{\delpm}{\delta\left(\E-\Ep \pm \hbar\omega\right)}
\newcommand{\delko}{\delta\left(\E-E_\text{F}\right)}
\newcommand{\delkp}{\delta\left(\Ep-E_\text{F}\right)}

\title{Phonon limited thermoelectric transport in Pb}
%
\author{F. Rittweger}
\address{Institut f\"{u}r Physik, Martin-Luther-Universit\"{a}t Halle-Wittenberg, DE-06099 Halle, Germany}
\ead{florian.rittweger@physik.uni-halle.de}
\author{N. F. Hinsche}
\address{Department of Physics, Technical University of Denmark, DK-2800 Kgs. Lyngby, Denmark}
\author{I. Mertig}
\address{Institut f\"{u}r Physik, Martin-Luther-Universit\"{a}t Halle-Wittenberg, DE-06099 Halle, Germany}
\address{Max-Planck-Institut f\"{u}r Mikrostrukturphysik, Weinberg 2, DE-06120 Halle, Germany}

\date{\today}

\begin{abstract}
We present a fully \textit{ab initio} based scheme to compute thermoelectric transport properties, i.e. the electrical conductivity $\sigma$ and thermopower $S$, in the presence of electron-phonon interaction. We explicitly investigate the \kdep structure of the \ESF, the coupling strength, the linewidth and the relaxation time $\tau$. We obtain a state-dependent $\tau$ and show its necessity to reproduce the increased thermopower for temperatures below the \textsc{Debye} temperature, without accounting for the phonon-drag effect. Despite the detailed investigations of various $\mathbf{k}$ and $\mathbf{q}$ dependencies, the presented scheme can be easily applied to more complicated systems.
\end{abstract}

\submitto{\JPCM}

\maketitle

\section{Introduction}
The interaction between fermions and bosons plays an important role in solid state physics and gives rise to many fundamental phenomena. While conventional phonon-mediated superconductivity is one of the most studied aspects in many-particle physics, recently the impact of electron-phonon (EP) coupling on carrier lifetimes \cite{Bernardi:2014,Park:2014,Liu:2017jf}, band renormalization \cite{Bostwick:2007be,Park:2007kx}, phonon-assisted absorption \cite{Noffsinger:2012va,Patrick:2014dp}, relaxation dynamics of excited carriers \cite{Brown:2016wa}, electronic mobilites \cite{Kaasbjerg:2012uy,Gunst:2016} and thermolectric properties \cite{Xu:2014uh,Liao:2015ca} got increased focus.
The demand for a consistent theoretical description of these phenomena led to many methodological improvements in recent years \cite{Giustino:2016vf,Ponce:2016dv} and enables computationally feasible ways to calculate and predict electron-phonon mediated material properties. 
As the archetype strong-coupling superconductor, Pb has been the material of choice in various theoretical studies of electron-phonon coupling properties \cite{Sklyadneva:2013et, Sklyadneva:2012, Heid:2010}. 
Within this manuscript we introduce an in-depth study of the electron-phonon interaction in fcc bulk Pb. Introducing our method based on density 
functional perturbation theory and improved tetrahedron interpolations we discuss anisotropic $\bm{k}$ and $\bm{q}$-dependent material properties, 
e.g. electron-phonon coupling parameter, linewidths and lifetimes as well as state-dependent band renormalizations. Based on the evaluation 
of the complex electron-phonon self energy we solve the Boltzmann transport equation in relaxation time approximation and elaborate details on the temperature-dependence 
of the electrical conductivity and thermopower within different approximations to the scattering rates. 
An enhancement of the thermopower at low temperatures due to the state-dependency of the relaxation times could be unveiled. 
More general we find, that simple single-sheeted \Fermi surfaces will yield smaller electron-phonon-coupling, i.e. larger relaxation times, compared to complicated multi-sheeted ones. 
Our in-depth analysis of an electron-phonon driven material, such as Pb, might help to gain a more systematical understanding of how to 
predict material properties influenced by electron-phonon-coupling.
\section{Methodology}

The electron and phonon properties were calculated within the framework of density functional perturbation theory as implemented in the Quantum \texttt{Espresso} package \cite{Giannozzi:2009}. We investigate Pb with the lattice constant 
$a_{lat}=9.2\ a.u.$ and use a scalar-relativistic norm-conserving pseudopotential with the Perdew-Zunger exchange correlation functional and the local density approximation\footnote{We note that Pb should be treated in a fully-relativistic description due to its large spin-orbit coupling but the methodical framework we present is not restricted to a special type of pseudopotential and relativistic effects can be added. The impact of these effects on properties determined by electron-phonon (el-ph) interaction is discussed in several papers \cite{DalCorso:2008,Sklyadneva:2012} and do not change our results qualitatively.}. 
The evaluation of the electron-phonon interaction is based on the el-ph matrix elements, which are first calculated on a coarse 18x18x18 k-grid and 8x8x8 q-grid and afterwards wannier-interpolated to finer grids using Wannier projections as implemented within the \texttt{EPW} code \cite{Ponce:2016dv}. The interpolation at each $\mathbf{k}$ point is performed with at least 200'000 $\mathbf{q}$ points.
 
The effectiveness of phonons with energy $\hbar \omega$ to scatter electrons 
can be expressed by means of the, \textit{a priori} state-dependent, \textsc{\'Eliashberg} spectral function within the quasi-elastic assumption 
\begin{equation}
\alpha^{2}F_{\ko}(\omega,\E)=\frac{1}{N_\mathbf{q}}\sum_{\q m}\delta(\omega-\omega_{\q})\left|\me\right|^2\delQE\quad,
\label{eq:a2Fk}
\end{equation}
where $\omega_{\q}$ is the phonon frequency of state $\mathbf{q}$ in mode $\nu$, $N_\mathbf{q}$ is the number of $\mathbf{q}$ points, $\E$ is the 
energy of an electron in initial state $\mathbf{k}$ and $n$-th band and $\mathbf{k}^\prime$ being the final state given by $\mathbf{k}+\mathbf{q}$. 
The kernel of the \ESF are the electron-phonon matrix elements 
\begin{equation}\label{eq:me}
\me=\sqrt{\frac{\hbar}{2NM\omega_{\q}}}\langle\Psi_{\ko+\q}\left|\partial V_{\q}\right|\Psi_{\ko}\rangle
\end{equation}
which manifest the interaction of the initial ($\Psi_{\ko}$) and final ($\Psi_{\ko+\q}$) electronic states via a phonon $(\q,\nu)$. Within $\partial V_{\q}$ is the first-order derivative of the Kohn-Sham potential with respect to the atomic displacements induced by a phonon 
mode $\nu$ with frequency $\omega_{\q}$.

Integration of the \textsc{\'Eliashberg} spectral function then directly leads to the electron-phonon coupling constant  
\begin{equation}
\lambda=2\int_0^{\infty} d\omega\,\frac{\alpha^2F(\omega)}{\omega}\qquad.
\label{eq:l}
\end{equation} 
and the phonon-induced electron linewidth  
\begin{eqnarray}
\label{eq:lw}
\Gamma_{\ko}(\E)&=& 2\pi\int_0^{\infty}d\omega\,\alpha^{2}F_{\ko}(\omega,\E)\mathcal{H}(b,f)\quad\text{with}\\
\mathcal{H}(b,f)&=&\left[1+2b(\omega)+f(\E+\hbar\omega)-f(\E-\hbar\omega)\right]\nonumber
\end{eqnarray}
and $b$ and $f$ being the \textsc{Bose-Einstein} and \textsc{Fermi-Dirac} distribution functions, respectively. 
The inverse linewidth is proportional to the el-ph relaxation time $\tau_{\ko}=\nicefrac{\hbar}{\Gamma_{\ko}}$. We note, that the corresponding electron-induced phonon linewidths $\Gamma_{\q}(\omega)$ can be easily obtained by summing over $\bm{k}$ instead of $\bm{q}$ in \eq{eq:a2Fk}.

An enhanced interest of the influence of el-ph interactions on $\mathbf{k}$- and energy-dependent electron, $\mathbf{q}$-dependent phonon properties as well as transport properties, especially in 2D-materials \cite{Hwang:2008,Park:2014,Liao:2015_1,Liao:2015_2,Gunst:2016} and bulk semi-conductors \cite{Piscanec:2004,Bernardi:2014,Tandon:2015,Li:2015}, came up recently. 
A quantity of major interest in transport calculations is the transport relaxation time $\tautr_{\ko}$. 
It can be calculated by adding an efficiency factor
\begin{equation}
1-\frac{\mathbf{v}_{\ko}\mathbf{v}_{\kp}}{\left|\mathbf{v}_{\ko}\right|\left|\mathbf{v}_{\kp}\right|}
\label{eq:velo_cos}
\end{equation}
to eq.(\ref{eq:a2Fk}), which accounts for the \textit{scattering in} term in the iterative solution of the Boltzmann transport equation and appears like a change of the absolute value of the velocity during the scattering process. The denominator is usually approximated by $\left|\mathbf{v}_{\ko}\right|^2$. This is a reasonable assumption in simple metals like lithium or sodium possessing isotropic, spherical \Fermi surfaces but does not hold for lead or even more sophisticated systems, which provide complex and anisotropic \Fermi surfaces. 

An equivalent way of calculating the relaxation time is given by solving a linearized \Boltzmann equation. 
While we already noted that within an iterative solution of the latter, the \textit{scattering in} term is here approximated by the by the efficiency factor \Eq{eq:velo_cos}, accounting for the \textit{scattering out} term only, would correspond to the relaxation time approximation (RTA). 
Recently, results solving the full \Boltzmann equation were published \cite{Li:2015,Fiorentini:2016ki}. Within it is shown that applying the full iterative solution accounting for el-ph scattering instead of a relaxation time approximation or an approximation including the efficiency factor, does not change the result noticeably for several studied metals and semiconductors. Nevertheless, this might not be valid for every system. 
Within the manuscript, the transport quantities $\tau^{\text{tr}}$, $\lambda^{\text{tr}}$ and $\left( \alpha^{2}F_\mathbf{k} \right)^{\text{tr}}$ explicitly account for the for the \textit{scattering in} term via the efficiency factor, while the latter is neglected for the spectroscopic counterparts $\tau$, $\lambda$ and $\left( \alpha^{2}F_\mathbf{k} \right)$.
 
Besides the linewidth and relaxation time, which are linked to the imaginary part of the complex el-ph self-energy $\Sigma(\E,T)=\Sigma^\prime(\E,T)+\mathfrak{i}\Sigma^{\prime\prime}(\E,T)$ via $\Gamma_{\ko}(T)=2\Sigma^{\prime\prime}_{\ko}(T)$, the renormalization of the electronic states, being mainly connected  to the real part of $\Sigma$, is less investigated in transport calculations. However, it may play an important role at low temperatures due to its large influence on the bandstructure around the \textsc{Fermi} energy. The renormalized energy $\Erenorm$ of the state $\kk$, as maximum of the spectral function $A(\mathbf{k},T)$, is given by
\begin{equation}
\Erenorm=\E-\Sigma^{\prime}(\Erenorm,T)\qquad.
\label{eq:renorm}
\end{equation}
Practically, one uses a modified \textsc{Kramers}-\textsc{Kronig} relation to obtain the real part of the self-energy from the imaginary part \cite{Levy_book:2000}.

\section{Results}

\begin{figure}[th]
\centering
\includegraphics[width=0.9\columnwidth]{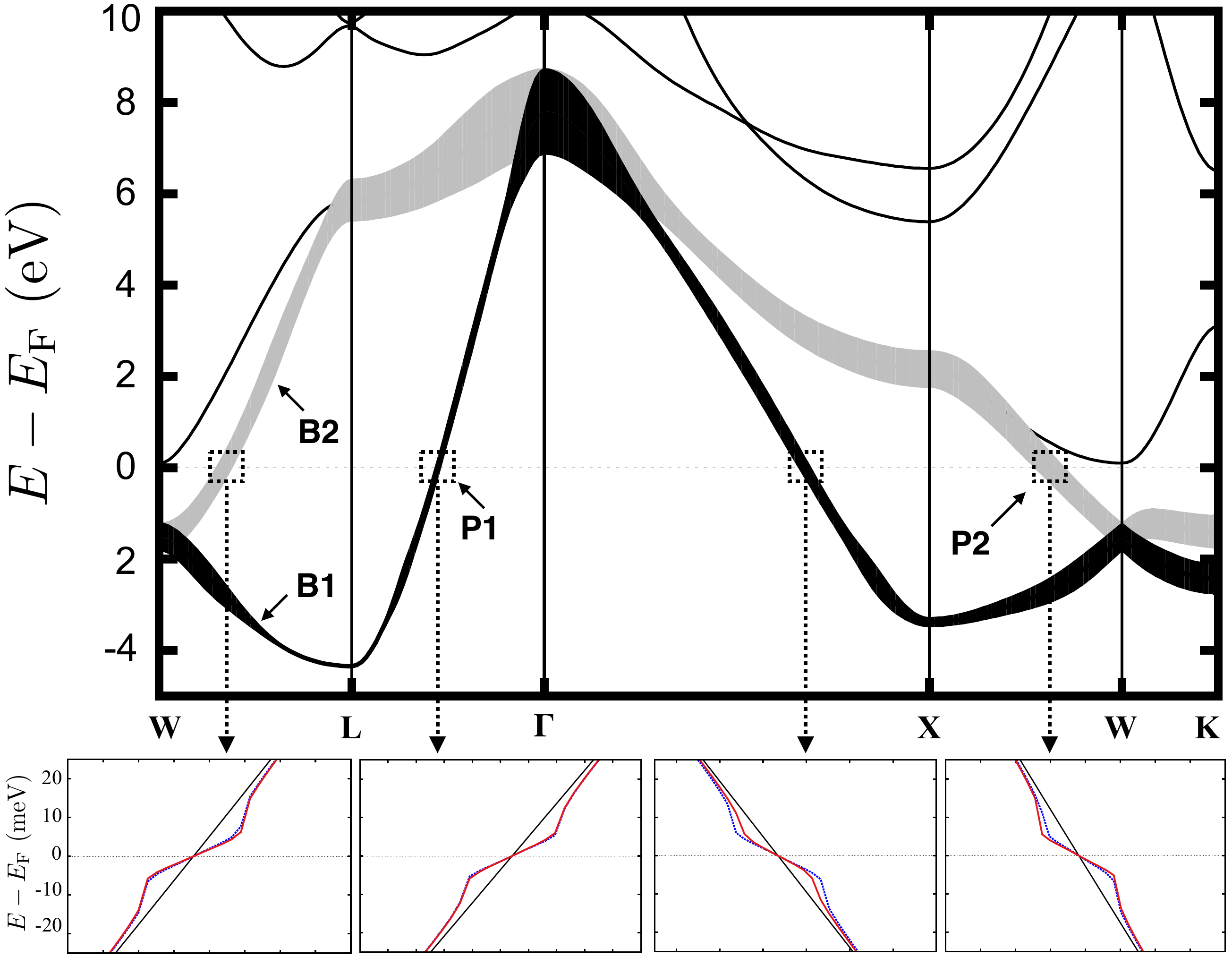}
  \caption{Calculated band structure of Pb within the scalar-relativistic framework. The width of the superimposed shaded areas onto the electronic states of band B1 and B2 is related to the magnitude of the \kdep linewidth at 300 K shown in figure \ref{fig3}(b). The values for the states P1 and P2 are 211 meV and 304 meV. The insets show the renormalized bandstructures around the \textsc{Fermi} energy at 10 K. The renormalization is done with the isotropic $\kk$-independent (dashed blue line) and the \kdep \ESF (solid red line). The \textit{kinks} appear around the maximal phonon energy.}
\label{fig1}
\end{figure}

The scalar-relativistic band structure of Pb shown in figure \ref{fig1} resembles previous calculations\cite{Zdetsis:1980,Heid:2010,Sklyadneva:2012}. Deviations from experiments occur due to spin-orbit interaction and are found primarily at the crossing points at W, L, $\Gamma$, on the high-symmetry line $\overline{\text{XW}}$ leading to avoided crossings. Since these regions are not close to the \textsc{Fermi} energy the influence on transport properties, like electrical conductivity or thermopower, is rather small at moderate temperatures.

\begin{figure*}[!ht]
\centering
\includegraphics[width=0.95\textwidth]{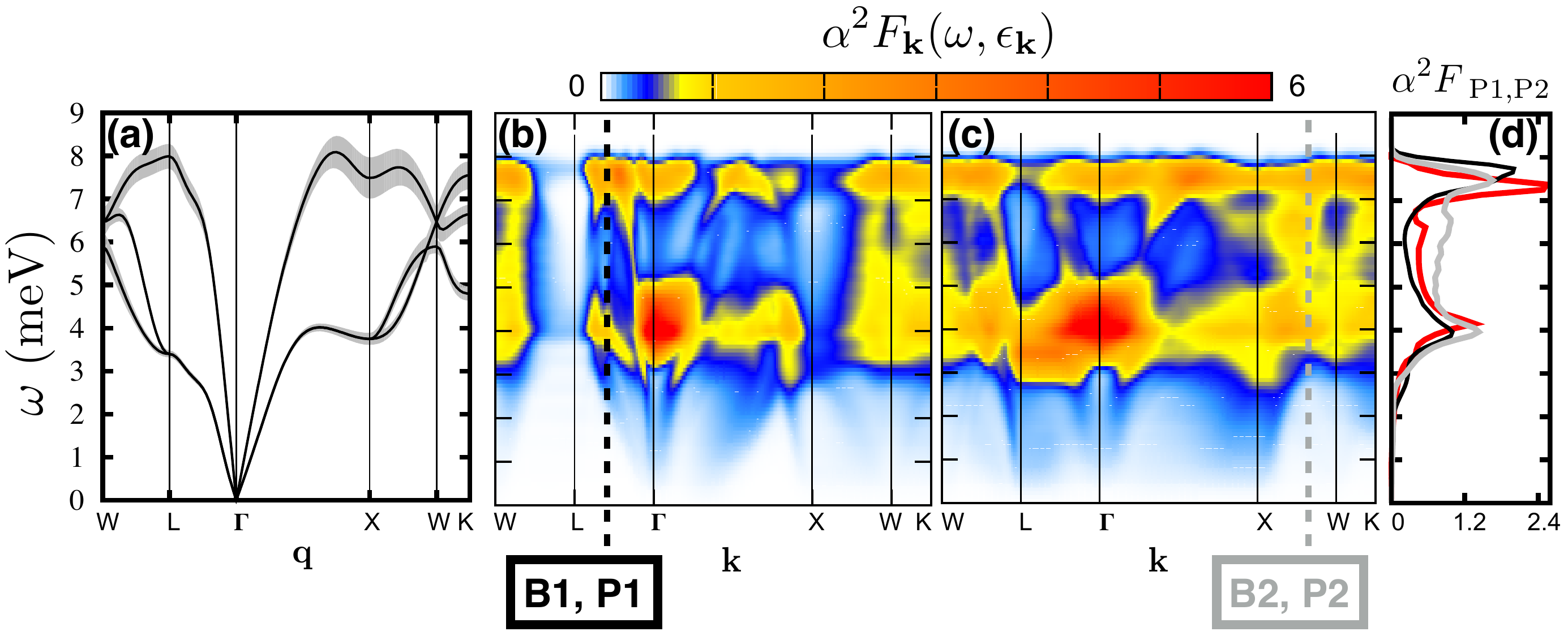}
  \caption{(a) Phonon dispersion curves with superimposed linewidths due to el-ph interaction. The maximum width corresponds to $\sim32\unit{meV}$. (b), (c) \kdep \textsc{\'Eliashberg} spectral function $\alpha^{2}F_\mathbf{k}(\omega,\epsilon_\mathbf{k})$ 
  along the high-symmetry line given in figure \ref{fig1} for the bands B1 and B2. The 
  colour-code is related to the values of $\alpha^{2}F_\mathbf{k}$. The states P1 and P2 are highlighted as dashed black and grey lines. (d) $\alpha^{2}F_\mathbf{k}$ of state P1 (black) and P2 (grey) on the Fermi surface. The contributions to the spectral function in P1 arise almost entirely from high-energy phonons, while the phonons involved in the coupling at P2 show a wider energy distribution. The red line shows the \Fermi surface averaged spectral function usually used when evaluating spectral and transport properties.}
\label{fig2}
\end{figure*}

The effect of the el-ph interaction onto the bare single-particle states is shown as superimposed linewidth and renormalization (inset fig. \ref{fig1}). The scattering of an electron with a phonon leads to a finite lifetime and linewidth of the electronic state. The absolute value of the linewdith depends on the interaction strength and the possible scattering phase space. The latter is determined by an energy conservation, $\Ep=\E\pm\wq$, and momentum conservation, $\kps=\kos\pm\qs$. Generally the phase space can be rather small in terms of the electronic energy scale, as the highest phonon energies are typically in the order of 10-100 meV, depending on whether metals \cite{Savrasov:1996} or semiconductors \cite{Giannozzi:1991} are in focus. Therefore, sampling the \Brillioun zone sufficiently is demanding. Consequently, it is complemented by the powerful Wannier interpolation scheme.
The renormalization of an electronic state due to el-ph interaction results in an effective gain or loss of the electron energy, related to its position relative to $\Ef$. The renormalization shown in the insets is done with the ordinary $\mathbf{k}$-independent \ESF, where we assumed an isotropic renormalization for all electronic states (dashed blue line), and the \kdep spectral function (solid red line). In both cases characteristic \textit{kinks} appear around $E=\Ef\pm\omega_\text{max}$, with $\omega_\text{max}\approx8\unit{meV}$ being the largest phonon frequency in lead. This leads to reduced velocities and an increased density of states (DOS) inside the window and to enhanced velocities and lowered DOS outside. While the isotropic approximation is resonable at certain $\kk$ points (around P1), it usually under- or overestimates the renormalization effect (around the crossing along $\overline{\Gamma\text{X}}$ and around P2). With increasing temperature the \textit{kinks} are disappearing and the influence on the states is negligible. For electronic states far away from $\Ef$, the possibility to absorb or emit a phonon is small and there is no significant influence of the el-ph interaction onto the electronic bands anymore no matter how small or large the temperature is. 

Figure \ref{fig2}(a) shows the calculated phonon dispersion of bulk Pb. A comparison with experimental data and other calculations including spin-orbit interaction, reveals differences to the scalar-relativistic calculation \cite{DalCorso:2008,Heid:2010}. Basically three features are missing. First, mode softening at certain points in the \Brillioun zone is not observed, most pronounced at X. Second, the absolute value of the phonon bandwidth is $\approx10\%$ to large. Third, \textsc{Kohn} anomalies, e.g. along the $\overline{\Gamma\text{K}}$-line, due to \Fermi surface nesting are not accounted for. Overall these features result in an enlarged el-ph coupling parameter and thus the relaxation times presented in this work will slightly overestimate possible experimental findings. 
Superimposed onto each mode is the phonon linewidth
\begin{equation}
\gamma_{\q}=2\pi\omega_{\q}\sum_{\mathbf{k}mn}\mes\delkp\delko\quad.
\end{equation}
It can be seen that the linewidth is large for high-energy phonons and drops to zero with $\omega\rightarrow0$. The renormalization of the phonon modes is already included within density functional perturbation theory (DFPT) since the real part of the phonon self-energy in the static limit equals the contribution to the dynamical matrix, which arises from the variation of the electron density. Hence, calculating phonon frequencies within DFPT already includes the renormalization due to the real part of the el-ph selfenergy.

The \kdep \ESF of the electronic states along the high-symmetry line given in fig. \ref{fig1} is shown in fig. \ref{fig2}(b) and (c) for bands B1 and B2. These quantities serve as direct input for the calculation of the relaxation times. First of all, the shape of each $\aaFs$ is quite similar compared to the isotropic \Fermi surface averaged \textsc{\'Eliashberg} function in Pb (red line in fig. \ref{fig2}(d)). Two peaks can be seen, more or less distinct from each other. One originates from high-energy phonons with approximately 7-8 meV and the other one arises from non-dispersive phonons around 3-4 meV. This pattern can be observed even at the L point in the band B1, where the scattering phase space is almost vanishing. The highest values of $\aaFs$ are found around $\Gamma$. Here, the electronic bandstructure favours the coupling to phonons with wavevector $\mathbf{q}$ along the $\overline{\Gamma\text{X}}$ direction. Since the dispersion of these phonons is rather flat nearby 3-4 meV, the contribution to $\aaFs$ is large. 
The states P1 and P2, lying on the \textsc{Fermi} surface are highlighted in fig. \ref{fig2}(c) and (d) with dashed black and grey lines and are displayed separately in figure \ref{fig2}(d). As mentioned before, both \textsc{\'Eliashberg} functions provide the two-peak-structure. While $\alpha^{2}F_\text{P1}$ is dominated by the low-dispersive mid- and high-energy phonons, the contributions to $\alpha^{2}F_\text{P2}$ are distributed over the energy range. Here, phonons are more dispersive and have larger velocities, heavily impacting the \textsc{\'Eliashberg} function. In both cases the coupling to phonons with an energy less than 2 meV is weak. These properties of the points P1 and P2 are characteristic for the anisotropy of the \ESF, the coupling constant, the linewidth and the relaxation time on the \Fermi surface discussed within this manuscript. 

\begin{figure}[htb]
\centering
\includegraphics[width=0.9\columnwidth]{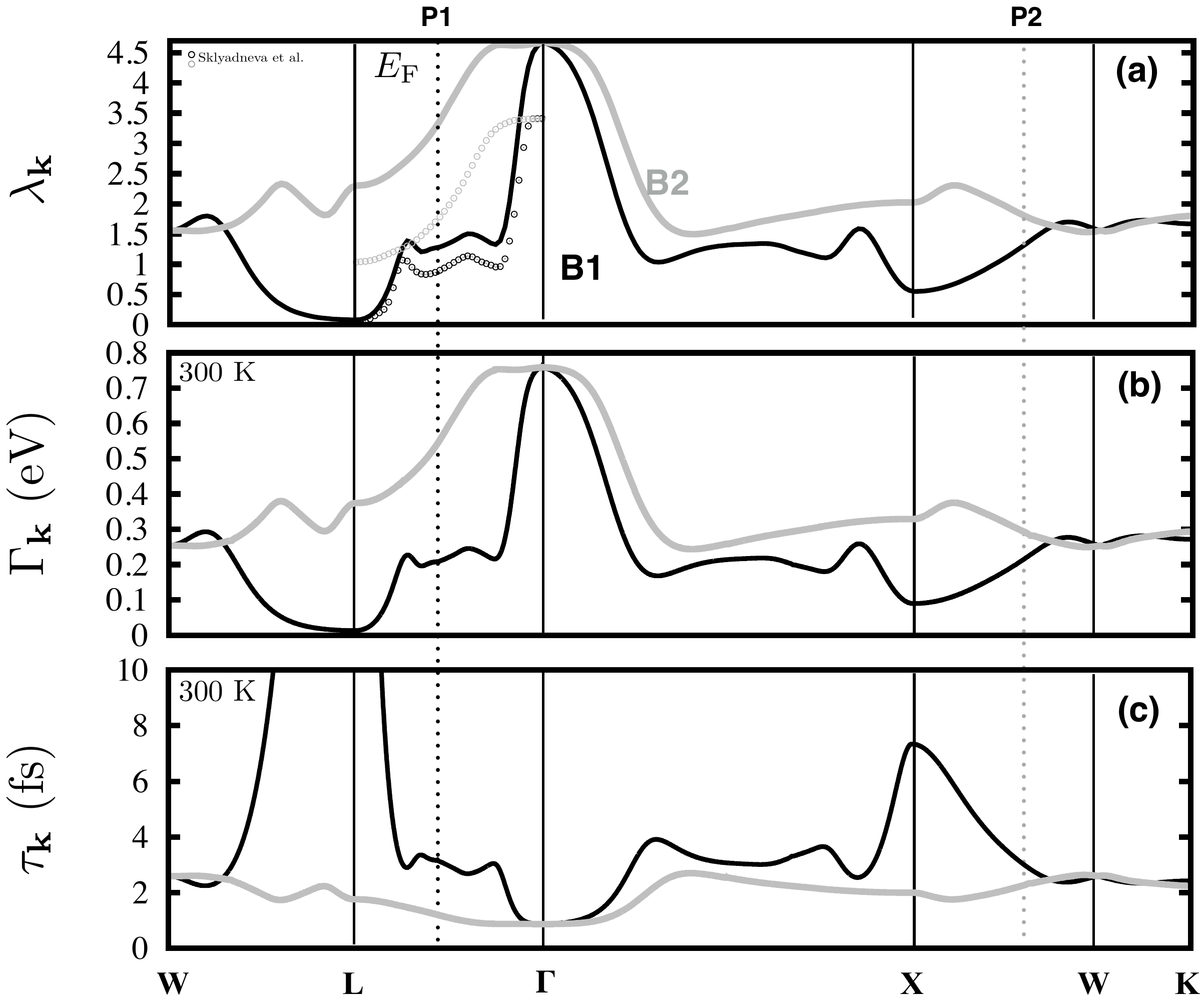}
  \caption{Various \kdep el-ph properties. The vertical dotted line indicate the crossing of the states P1 and P2 of the \textsc{Fermi} energy of the bands B1 (black) and B2 (grey). All quantities in (a)-(c) were calculated via an integration over the \textsc{\'Eliashberg} spectral function given in figure \ref{fig2}(b) and (c). (a) Electron-phonon coupling parameter. (b) Linewidth at 300K. (c) Spectral electron-phonon relaxation time obtained from the linewidths in (b). At 300K, $\tau_\mathbf{k}$ is about $3.15\unit{ fs}$ and $2.15\unit{ fs}$ in P1 and P2, respectively.}
\label{fig3}
\end{figure}

A certain quantity obtained from the \textsc{\'Eliashberg} function is the \kdep coupling constant $\lambda_\mathbf{k}$, which is shown in fig. \ref{fig3}(a). The calculated values along the $\overline{\Gamma\text{L}}$ line are qualitatively in good agreement with data from Sklyadneva \textit{et al.} \cite{Sklyadneva:2012}. They show the same k-dependency for B1 as well as for B2. 
The coupling constants $\lambda$ for P1 and P2 are 1.30 and 1.79. The \Fermi surface averaged coupling strength is 1.35, which is higher than reported by Sklyadneva \textit{et al.} but still in good agreement with other published results based on a scalar-relativistic treatment of lead \cite{Liu:1996,Floris:2007,Aynajian:2008}. 
At higher temperatures, the factor $\left[1+2b(\omega)+f(\E+\omega)-f(\E-\omega)\right]$ in eq. (\ref{eq:lw}) is almost determined by $\nicefrac{1}{\omega}$ and therefore the shape of the linewidth (fig. \ref{fig3}(b)) and the coupling strength are quite similiar to each other. $\Gamma_\mathbf{k}$ is 0.211 eV and 0.295 eV for P1 and P2, respectively. The inverse linewidth gives rise to the relaxation time, shown in fig. \ref{fig3}(c). The largest value can be observed for B1 at L since the scattering phase space is small and nearly no coupling between electrons and phonons occurs. Conversely, $\tau$ exhibits the smallest value at $\Gamma$, because of the large coupling between electrons and phonons. The anisotropy of the \textsc{\'Eliashberg} function at the \Fermi energy yields relaxation times of 3.15 fs and 2.15 fs for P1 and P2. The \Fermi surface averaged relaxation time within the high-temperature limit, $\tau(E_\text{F})=\nicefrac{\hbar}{2\pi\text{k}_\text{B}\lambda T}$, at $T=300\unit{K}$, when the temperature is much higher than the \textsc{Debye} temperature, is 3.00 fs. Despite the larger $\tau$ at P1, scattering in the outer Fermi sheet (B2) is the dominant part (c.f. \F{fig5}(a)). The latter is related to the topology of the \Fermi surface and will be discussed subsequently. 

\begin{figure}[htb]
\centering
\includegraphics[width=0.9\columnwidth]{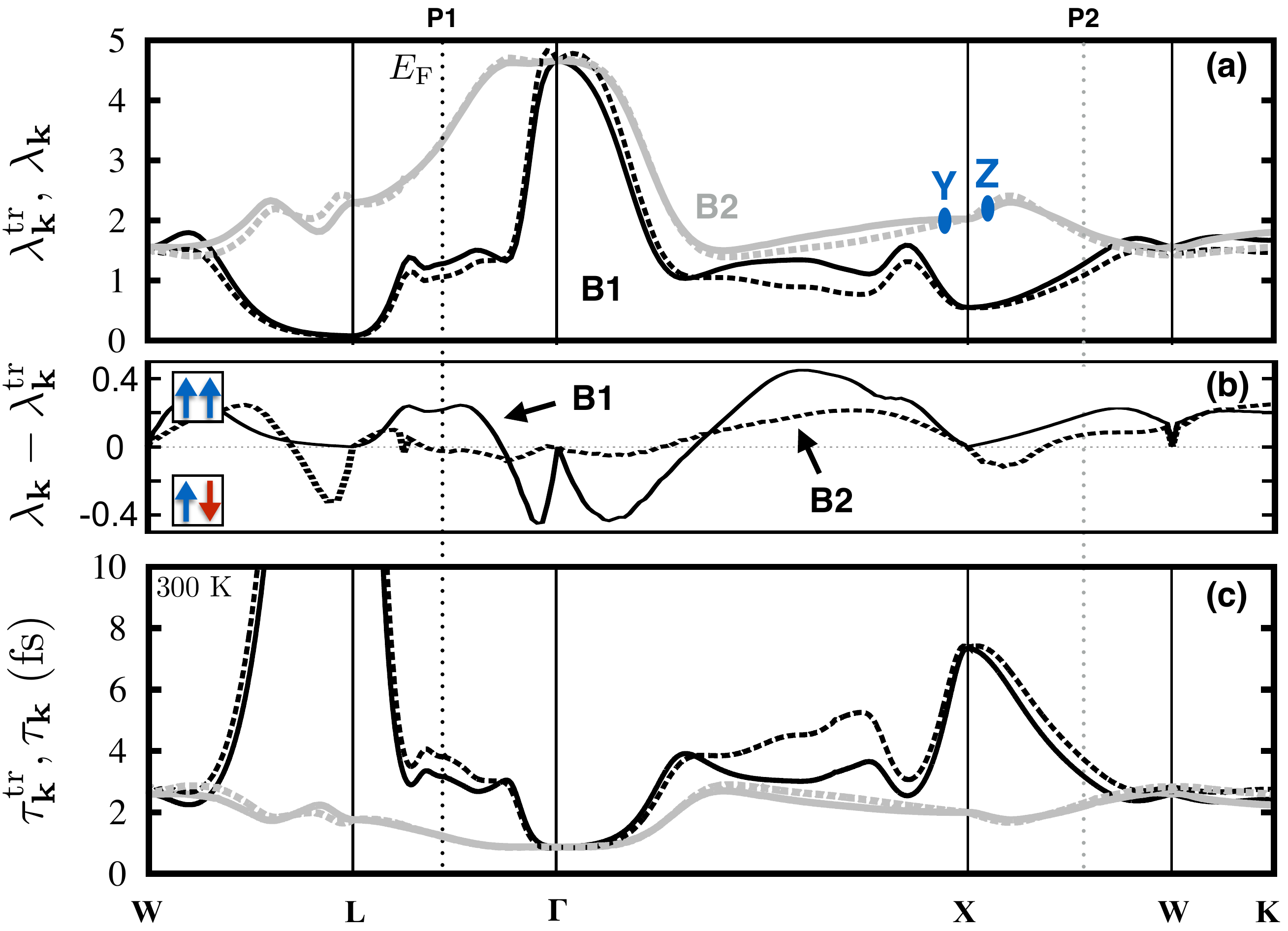}
  \caption{Comparison of the \kdep transport properties with its spectroscopical counterparts. The graphical layout is the same as in figure \ref{fig3}. (a) Coupling constants $\lambda_\mathbf{k}^\text{tr}$ (solid lines) and $\lambda_\mathbf{k}$ (dashed lines). Y and Z are highlighted k-points which are discussed in more detail within the main text. (b) Difference $\lambda_\mathbf{k}-\lambda_\mathbf{k}^\text{tr}$, at some $\mathbf{k}$ points either forward- (positive, $\uparrow\uparrow$) or backward-scattering (negative, $\uparrow\downarrow$) is favoured. (c) Relaxation times $\tau_\mathbf{k}^\text{tr}$ (solid lines) and $\tau_\mathbf{k}$ (dashed lines) at 300 K.}
\label{fig4}
\end{figure}

The spectroscopical properties discussed so far are of minor interest when calculating transport properties. For the latter one needs to include the efficiency factor given in eq. (\ref{eq:velo_cos}). 
The overall trend, obtained by the spectroscopical properties, does not change drastically when taking the efficiency factor into account (see fig. \ref{fig4}(a), (c)). Nevertheless, certain $\kk$ points favour either effective forward- ($\uparrow\uparrow$) or backward-scattering ($\uparrow\downarrow$), depending on the sign of the function $\lambda_\mathbf{k}-\lambda_\mathbf{k}^\text{tr}$, which is given in fig. \ref{fig4}(b). Which one is favoured is usually not predictable \textit{a priori} and depends on the geometry of the \Fermi surface, the coupling strength, or on a delicate mixture of both. A detailed analysis of the scattering events would be necessary and will be exemplary done at the $\kk$ points Y and Z of the band B2 around the high-symmetry point X. It turns out that the efficiency factor decreases (increases) $\aaFs$ at lower phonon energies at Y (Z) but keeps $\aaFs$ relatively uneffected at higher energies. Hence, the resulting transport coupling constant is smaller (larger) and forward (backward)-scattering is favoured. The decrease of $\lks$ due to the efficiency factor at Y is related to the geometry of the \Fermi surface only, which can be seen in a constant coupling approximation. This is not the case at Z, where the coupling strength itself is more important than the geometry of the \Fermi surface. Directly at the high-symmetry points, $\lktrs$ equals $\lks$ because of the vanishing velocity at the zone boundaries.

As mentioned before, on average the transport relaxation time differs only slightly from its spectroscopical counterpart. However the changes in {\bf{k}}-space can be substantially. $\tautr$ is about 3.81 fs at P1 and 2.32 fs at P2, which amounts to an increase of 21\% and 8\% compared to $\tau$.

\begin{figure*}[t]
\centering
\includegraphics[width=0.75\textwidth]{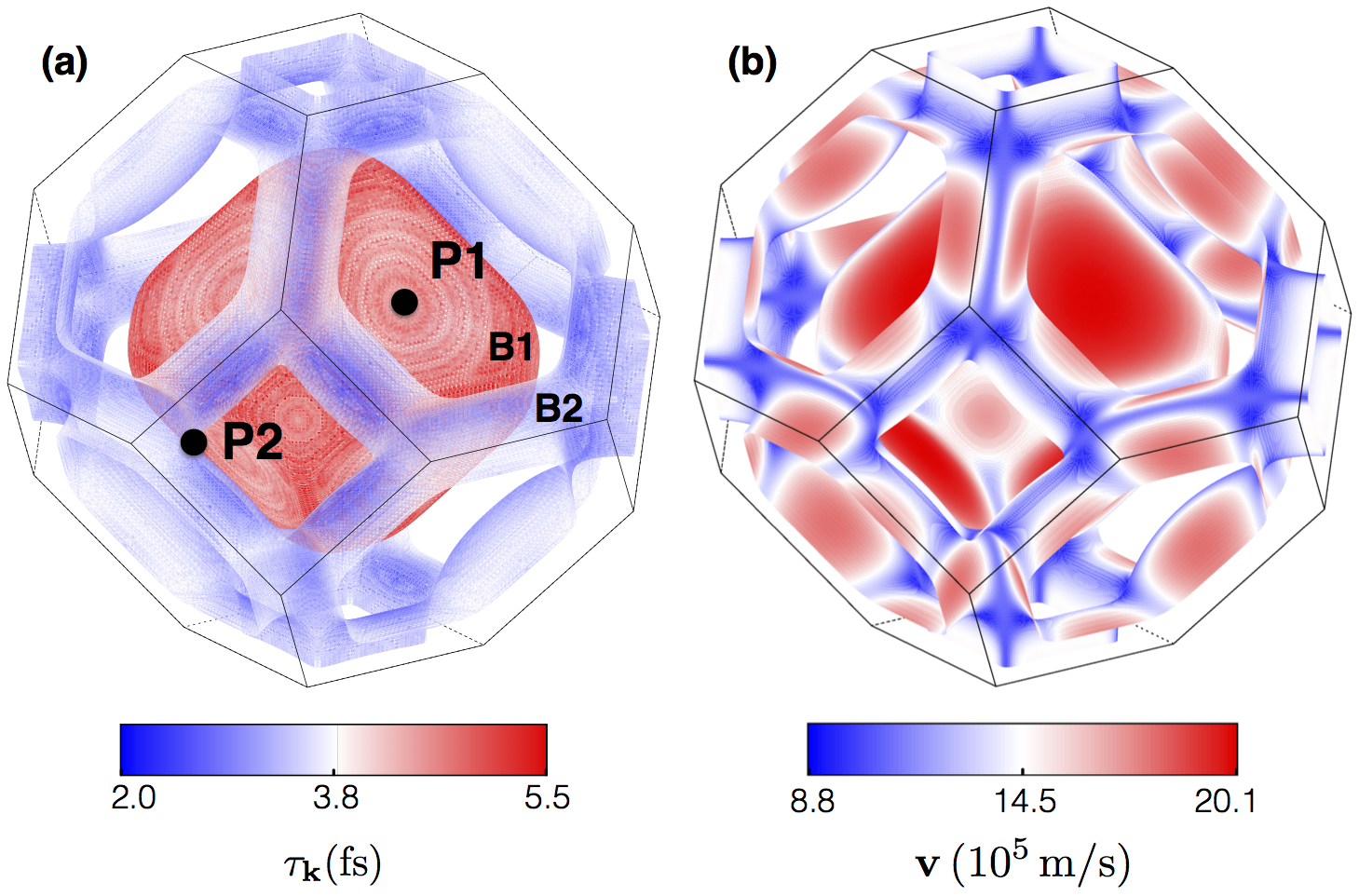}
  \caption{(a) \Fermi surface of Pb showing the \kdep electron-phonon relaxation times $\tau_\mathbf{k}$ at $T=300\unit{ K}$ superimposed onto the two sheets of the \Fermi surface. $\tau_\mathbf{k}$ is obtained by the solution of the \Boltzmann equation within the relaxation time approximation.
  The relaxation time is almost doubled in the inner sheet (B1) indicating a less efficient electron-phonon coupling compared to the outer sheet (B2). $\tau_\mathbf{k}$ in P1 and P2 is about $4.60\unit{ fs}$ and $3.27\unit{ fs}$, respectively. (b) \Fermi surface with the superimposed \Fermi velocities.}
\label{fig5}
\end{figure*}

\begin{figure*}[th!]
\centering
\includegraphics[width=0.95\textwidth]{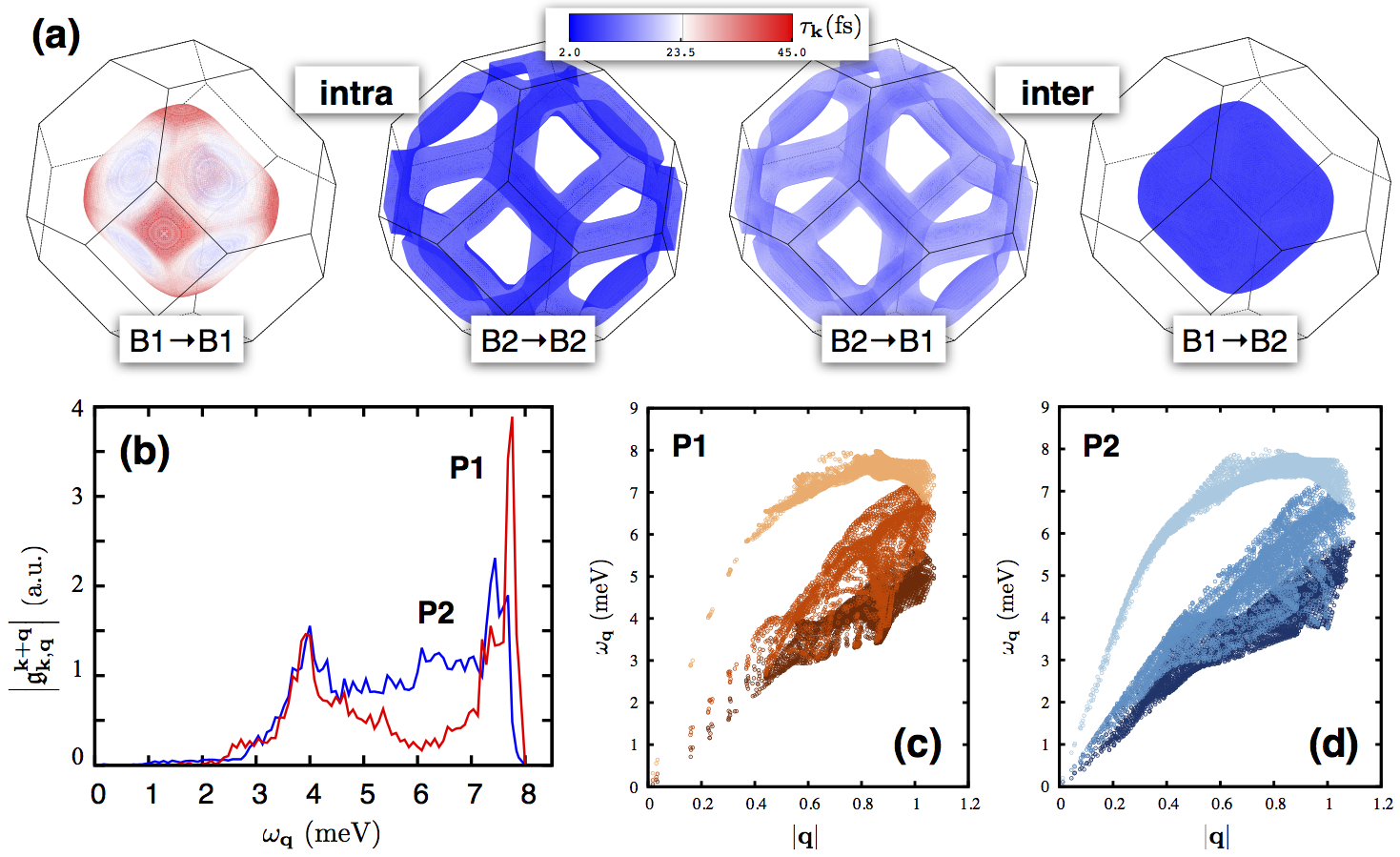}
  \caption{(a) Contributions to $\tau_\mathbf{k}$ arising from intra- and interband transitions. Shown are the initial states. For both transition types, scattering at/to the inner sheet (B1) results in larger relaxation times. The \Fermi surface averaged $\tau_\kk$ is dominated by the scattering at/to the outer sheet (B2). (b) Electron-phonon matrix elements plotted versus the corresponding phonon energies for P1 and P2 shown in fig. (\ref{fig5}). The latter one exhibits an uniform and in total larger electron-phonon coupling strength, which explains the smaller relaxation time. (c,d) Phonon energies versus absolute momentum for phonons being involved into the coupling from the initial states P1 and P2. Each color corresponds to one phonon mode.}
\label{fig6}
\end{figure*}

After the discussion of \kdep properties throughout the \Brillioun zone at arbitrary energies in the first part, we now want to look at the specific $\mathbf{k}\text{-resolved}$ structure of the relaxation time at the \Fermi energy.
Hence, we start from a dense mesh with more than 40'000 $\kk$ points in the irreducible part of the \Brillioun zone and extract $\kk$ points at the \Fermi surface with an adaptive tetrahedron method\cite{Zahn:2011}. We end up with a total number of 460'000 points at the whole surface, which are paired with a dense phonon mesh of 2.2 million $\qq$ points in the full \Brillioun zone to ensure convergence of the transport properties.

Figure \ref{fig5}(a) shows the \Fermi surface of Pb with the superimposed $\tau_\kk$. The relaxation time of the inner sheet (B1) is always larger than that of the outer sheet (B2) and hence the averaged relaxation time $\bar{\tau}$ of the inner sheet is ~30\% larger. The latter \textit{a priori} would make the usage of two constant relaxation times in transport calculations feasible. 
The overall structure of $\tau_\kk$ is anisotropic throughout the BZ, which can be related to the topology of the \Fermi surface itself. A certain initial state is coupled by different phonons to a whole set of final states, which is shown in fig. \ref{fig6}(c) and (d). Since the \Fermi surface is much more anisotropic around P2 than P1 and both bands are close to each other, the variety of the momenta of the coupled phonons is larger in the initial state P2. For example, only a few phonons with small wave vectors couple to P1, because its surrounding favours only coupling in a certain direction. Therefore, a simple \Fermi surface could partially act as a phonon filter.\footnote{Additional figures adding to this discussion can be found in the supplemental material.}
Besides the geometrical aspects, the $\mathbf{q}$-dependent coupling strength (fig. \ref{fig6}(a)) shows two distinct peaks for P1 and is more uniform for P2, which coincides with the \ESF (see fig. \ref{fig2}(d)). 
Decomposing the relaxation time into contributions from intra- and interband scattering, one can see that $\tau_{E_\text{F}}$ is dominated by transitions B1$\rightarrow$B2 and B2$\rightarrow$B2 (fig. \ref{fig6}(b)). Large relaxation times from transitions B1$\rightarrow$B1 and B2$\rightarrow$B1 are suppressed as one needs to add up the scattering rates of all transitions instead of the scattering times to account for \textsc{Matthiesen}'s rule. A comparison of the intraband transitions indicates that a simple single-sheeted \Fermi surface exhibits larger relaxation times than complicated multi-sheeted ones. In the case of Pb the enhancement compared to $\bar{\tau}$ is up to 600\% for the averaged values $\bar{\tau}^\text{intra}_\text{B1}$ and $\bar{\tau}^\text{intra}_\text{B2}$. 

We note, that there is probably no obvious qualitative correlation between the carrier velocity $\mathbf{v}_\kk$ of an initial state and its relaxation time $\tau_\kk$ as one might suspect from recent results found for noble metals by Mustafa \textit{et. al}\cite{Mustafa:2016}. The relaxation time of a free electron in the Drude model is given by
\begin{equation}
\tau_{D}=\frac{m^*}{n e^2\rho}\quad,
\end{equation}
with the resistivity $\rho$, carrier density $n$ and effective mass $m^*$. The latter is indirect proportional to the velocity $m^*\propto v^{-1}$, which transfers directly to the relaxation time yielding $\tau_\kk\propto(\mathbf{v}_\kk)^{-1}$ for a spherical Fermi surface. Surprisingly, this is partially valid in lead, since scattering from an initial state at the inner simple \Fermi sheet (B1) shows an indirect proportionality between the relaxation time and the velocity as well. On the other hand, $\tau_\kk$ is direct proportional to $\mathbf{v}_\kk$ for scattering starting at the outer \Fermi sheet (B2). Both figures are shown separately in the supplemental material. Additionally, an analysis of the orbital character of initial and final states, to predict smaller or larger relaxation times, as suggested in Ref.~\cite{Mustafa:2016}, times fails for the particular case of Pb as well since states around the Fermi energy are dominated by p-type orbital character.

\begin{figure}[htb]
\centering
\includegraphics[width=0.9\columnwidth]{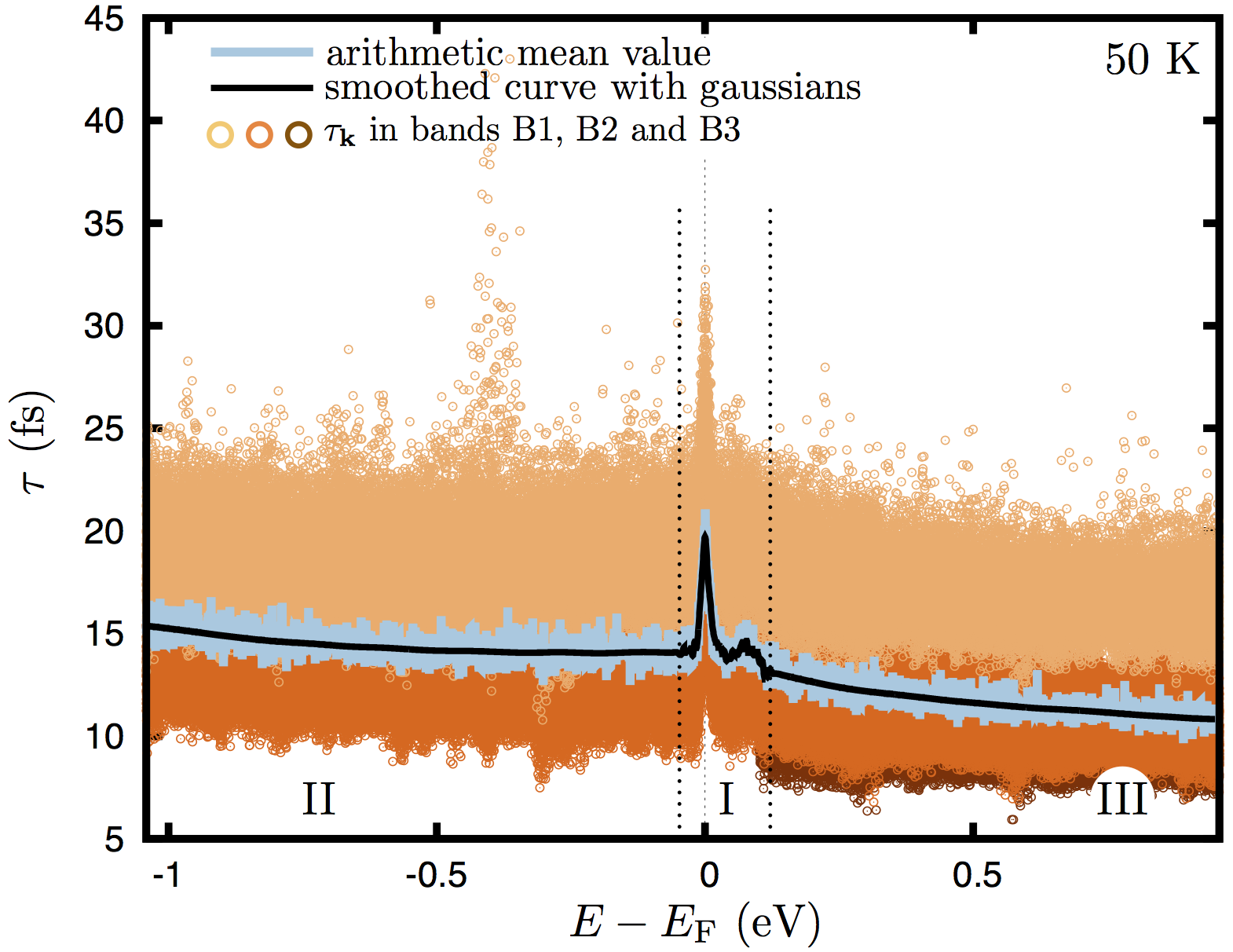}
  \caption{The state-dependent relaxation times $\tau_\kk$ for the bands B1 and B2 as well as the next higher band are shown (open circles). Starting from $\tau_\kk$, the energy-dependent $\tau$ is given calculated by an arithmetic mean value (light blue line) and an adaptive smearing method (black line). Further details are given in the text.}
\label{fig7}
\end{figure}

After the detailed investigation of relaxation times in single states $\kk$ at the \Fermi energy, the energy dependence of $\tau$ is investigated. The latter allows to perform accurate transport calculations, which are a step beyond the constant relaxation time approximation (CRTA), and which we will call energy relaxation time approximation (ERTA).

Figure \ref{fig7} demonstrates the approach we are using to obtain an energy-dependent relaxation time from a \kdep one. Each open circle corresponds to a relaxation time in state $\kk$ and related energy $\epsilon_\kk$. Since we are interested in $\tau(E)$, we compute the arithmetic mean value (AMV) of the scattering rates. The inverse quantitiy, $\tau_\text{AMV}(E)$ is shown as a solid light blue line. However, the result depends on the width of the energy interval and is rather jagged \footnote{Further details are shown in the supplemental material}. The latter does not affect the results of the electrical conductivity but strongly influences the calculation of the thermopower due to its sensitivity to the slope of the transport distribution function at the \Fermi energy. Therefore we use an adaptive smearing method to smoothen the arithmetic mean values (solid black line). The smearing is done with adaptive gaussians of certain widths, which are related to the slope of $\tau_\text{AMV}$. Here we distinguish between three segments. One includes the peak at the \Fermi energy and the signatures slightly above in energy (I) and the other two consist of the remaining parts below (II) and above (III) the first one \footnote{The used smearing values are listed in the supplemental material.}.

This scheme allows us to obtain a band-resolved energy-dependent relaxation time. Evidently one can see that the relaxation time of band B1 is larger compared to B2. This holds for every temperature at every energy because the magnitude of $\tau$ is mainly given by $\mathcal{H}(b,f)$. Additionally, the distribution functions $b$ and $f$ are responsible for the low-temperature peaks in $\tau(E)$ at $\Ef$. These peaks originate from the decreasing scattering phase space for phonons at temperatures below the \textsc{Debye} temperature ($\Theta_\text{D}\approx$ 95K), since phonons with longer wavevectors are frozen out and the number of scattering events is drastically reduced. 
The little bumb around 100 meV above $\Ef$ marks the bottom of the next higher band and the relaxation time decreases above this energy due to the additional electronic scattering channels. 

\begin{figure}[hbt]
\centering
\includegraphics[width=0.9\columnwidth]{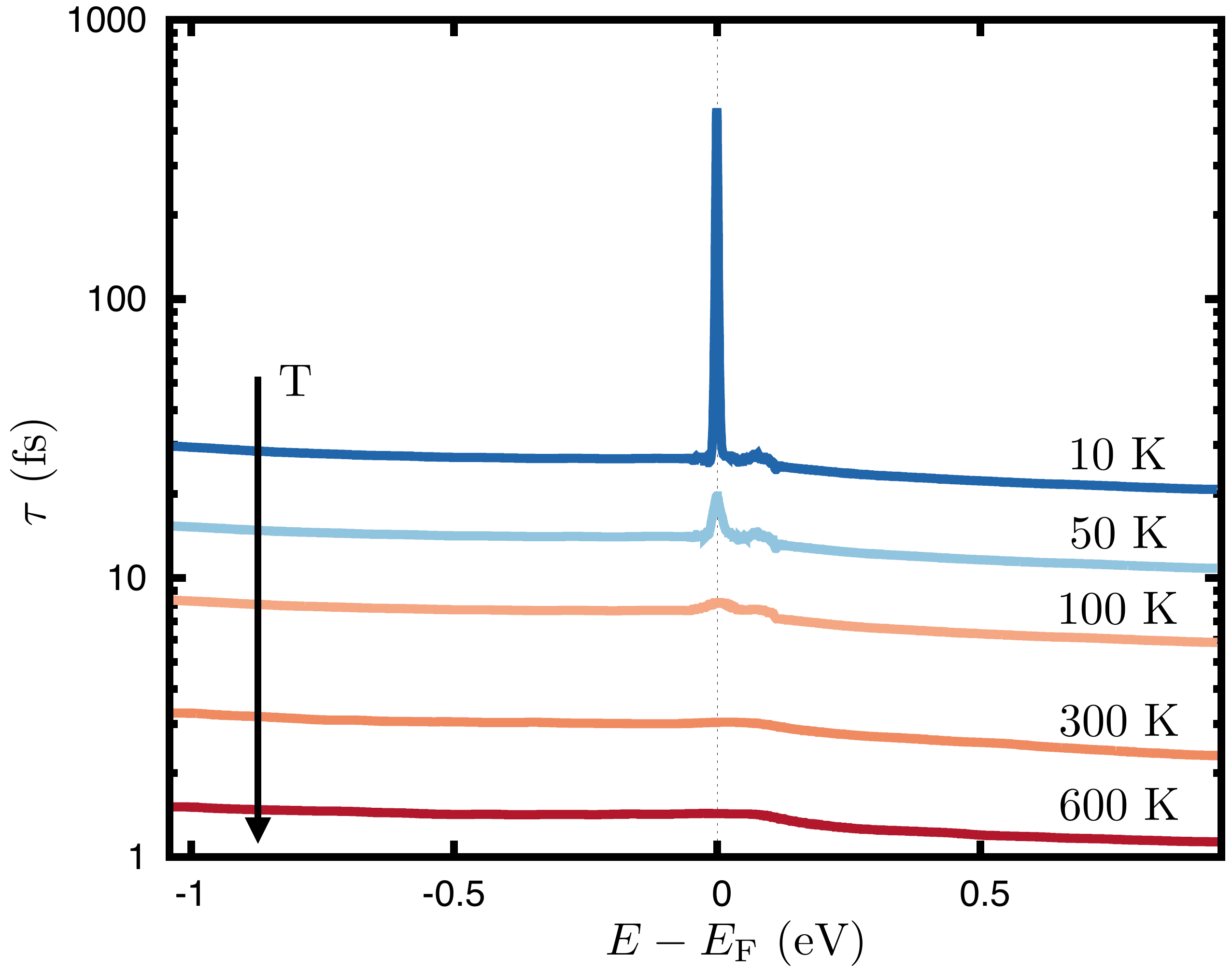}
  \caption{Energy-dependent relaxation times are shown for several temperatures from 10K (blue line) to 600K (red line). $\tau$ is decreasing with increasing temperature as indicated by the black arrow. The peak at $\Ef$ is characteristic for temperatures below $\Theta_\text{D}$ and originates from a freeze-out of long-wavevector phonons and a decreased scattering phase space.}
\label{fig8}
\end{figure}

The temperature has no significant impact on the functional behaviour of $\tau(E)$ apart from $\Ef$ and only determines its magnitude (see fig. \ref{fig8}). However, the shape of $\tau$ near the \Fermi energy strongly depends on $T$ as discussed before. Especially width and slope of the peak at $\Ef$ are important for the calculation of the thermopower. Having this in mind we will distinguish between the following approximations for the relaxation time. 
First of all we apply a relaxation time approximation, which is constant in energy but depends on temperature and can be either band-depended, bd-cRTA($T$), or not, cRTA($T$). The values are estimated from $\tau(E)$ at the \Fermi energy for each temperature separately. Next, we use a common approximation of $\tau$, including the density of states $\tau_\text{DOS}(E,T)=\nicefrac{A(T)}{\text{DOS}(E)}$. Here $A(T)$ is temperature-dependent. Finally, we calculate the transport properties with an energy- and temperature-dependent relaxation time based on a state dependency, as shown in fig. \ref{fig8}.

\begin{figure}[htb]
\centering
\includegraphics[width=0.9\columnwidth]{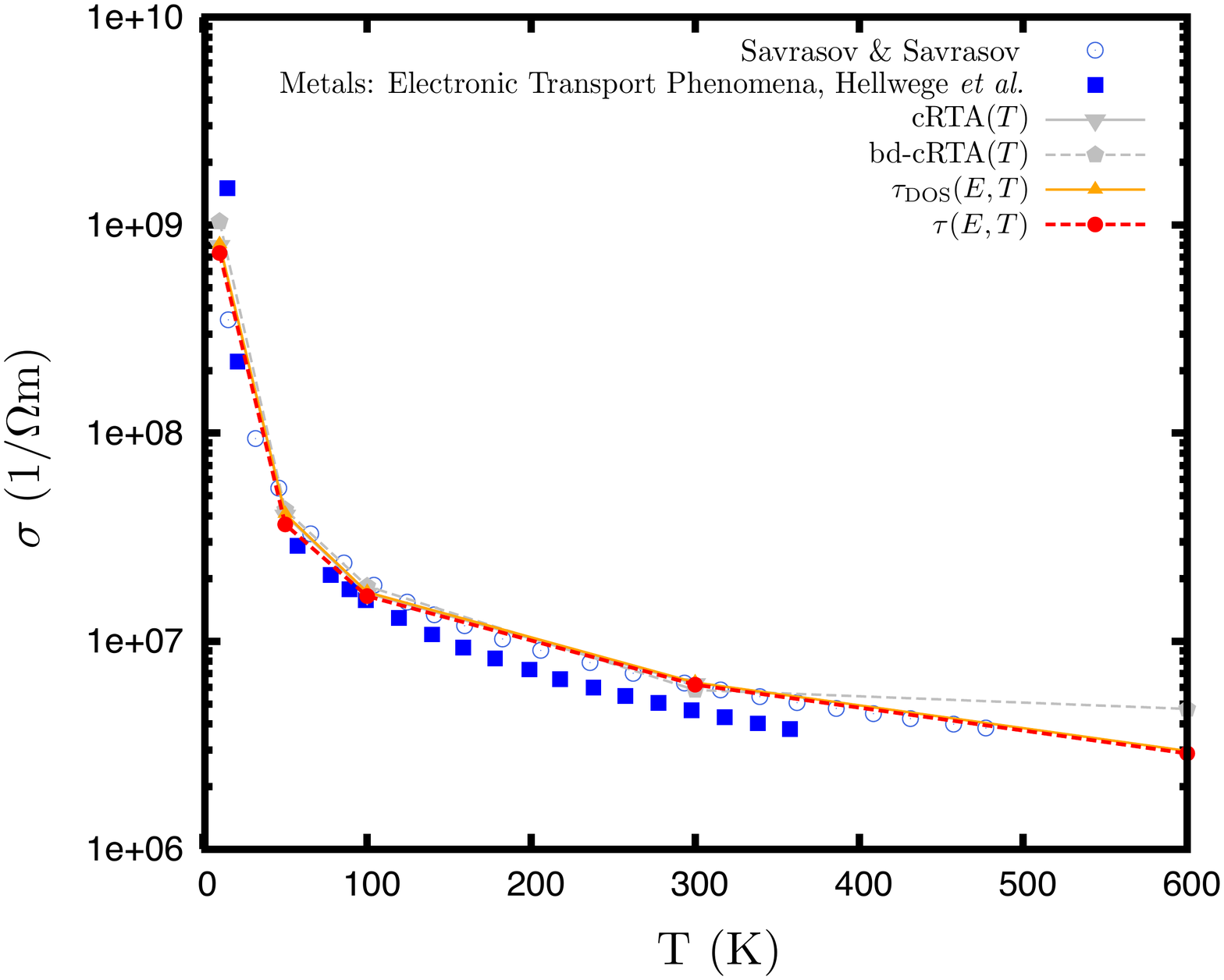}
  \caption{Calculated electrical conductivity of lead. The electron-phonon interaction is included within several approximations (lines with points). The results are compared to experimental \cite{Hellwege:1982:book} and  theoretical \cite{Savrasov:1996} data (closed squares and open circles).}
\label{fig9}
\end{figure}

We obtain the electrical conductivity and thermopower within the \textsc{Boltzmann} transport theory and the use of generalized transport coefficients
\begin{eqnarray}
& \mathcal{L}^{(n)}(\mu, T) = \\
& \int dE \; \Sigma(E) \cdot (E-\mu)^{n}\left( -\frac{\partial f^0(\mu,T)}{\partial E} \right)_{E_k=E}\quad,\nonumber
\label{eq:TcoeffE} 
\end{eqnarray}
where $\Sigma$ is the transport distribution function (TDF) introduced by \textsc{Mahan} and \textsc{Sofo}\cite{Sofo:1996} here explicitly accounting for the el-ph transport relaxation time $\tautr$.

The calculated electrical conductivities within the before mentioned relaxation time approximations agree very well with published experimental and theoretical values (fig. \ref{fig9}). 
While comparing the used approximations basically no difference can be found and even the temperature-dependent cRTA gives acceptable results. The reason for this is given by the integrand of eq. (\ref{eq:TcoeffE}). Here, the electrical conductivity is proportional to the zeroth moment, $\sigma(\mu,T)\approx\mathcal{L}^{(0)}$ and the important integration range is determined by the energy width of the derivative of the \textsc{Fermi}-\textsc{Dirac} distribution function, which is in the order of several  $k_\text{B}T$. Within the approximations for $\tau$ there is almost no difference in the integrand around $\Ef$. 

The situation is different for the diffusive thermopower, as $S$ is proportional to $\mathcal{L}^{(1)}(\mathcal{L}^{(0)})^{-1}$ and the specific functional behaviour of $\Sigma$ nearby $\Ef$ is crucial and changes whether one uses $\tau(T)$, $\tau_\text{DOS}(E,T)$ or $\tau(E,T)$. In the first two cases, the slope in the energy of the TDF does not change with temperature, only the absolute values are affected by $T$. The obtained thermopower within these approximations is almost linear in $T$ and can not reproduce the experimental data for $T<\Theta_\text{D}$ (see fig. \ref{fig8}, grey/orange solid and grey dashed lines). Nevertheless, they fit quite well at temperatures far above $\Theta_\text{D}$, which could be expected since all additional features in $\tau(E,T)$ around $\Ef$ arising from low temperatures are neglected anyway (see fig. \ref{fig7}) and approximations considering the whole phonon spectrum are reliable, as all phonon states are occupied. 

If we include the energy-dependent relaxation time $\tau(E,T)$ the enhanced thermopower at $T<\Theta_\text{D}$ can be very well described. Since we use the standard approach to solve the \Boltzmann equation with electron-phonon interaction, i.e. phonons are treated within equilibrium\cite{Ziman_book:1963}, the experimentally observed enhancement of $S$ is not or at least not solely caused by the phonon-drag effect. The last is usually the explanation for an increased thermopower in low-doped semiconductors but might only partially contribute to an enhanced thermopower in metals. In fact, the peaked asymmetric structure of $\tau(E,T)$ at low temperatures stemming from an explicit description of the electron-phonon interaction gives rise to an additional contribution to the thermopower even in the diffusive part, which then qualitatively reproduces the measured thermopower. These findings agree well with the qualitative description of an enhanced absolute thermopower $S_\text{el-ph} \propto S_0(1+\lambda)$ at low temperatures.

\begin{figure}[htb]
\centering
\includegraphics[width=0.9\columnwidth]{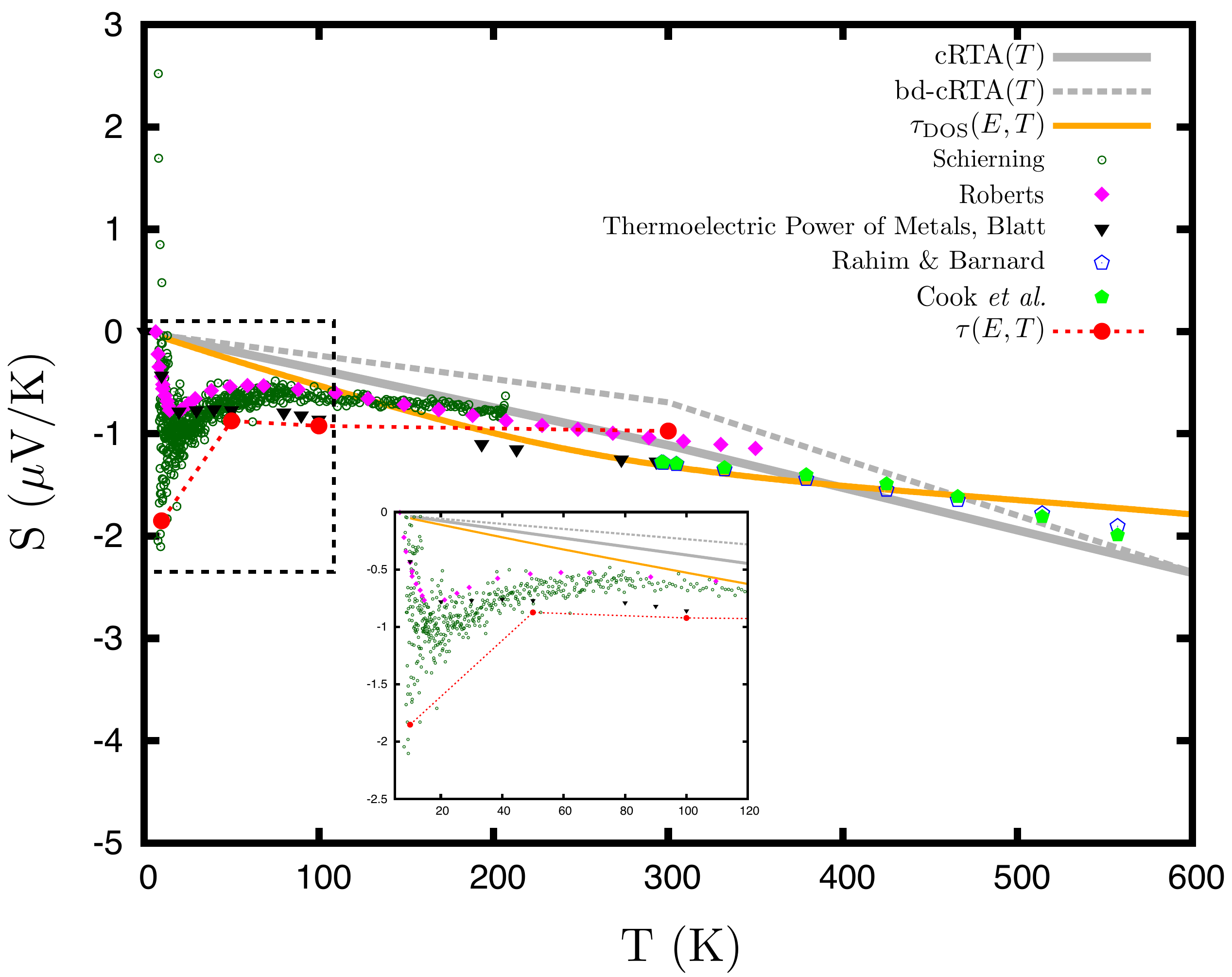}
  \caption{The calculated thermopower of lead including electron-phonon interaction within several approximations (lines and lines with points) is compared to a variety of measured values (different types of points) \cite{Schierning,Roberts,Blatt_book:1976,Rahim:1975,Cook:1974}. While the relaxation time approximations can not reproduce the experimental data at low temperatures, the energy- and temperature-dependent $\tau$ describes them very well.}
\label{fig10}
\end{figure}

\section{Summary}

We presented a scheme to calculate the influence of the electron-phonon interaction onto several transport relevant properties like coupling strength, linewidth and relaxation time based on \textit{ab initio} methods. We explicitly investigated the \kdep structure of $\tau$ at the \Fermi energy and explained its anisotropic character of bulk Pb. We obtained an energy-dependent relaxation times, which we are basis to calculate the electrical conductivity and thermopower. A comparison with various relaxation time approximations shows no significant difference for the electrical conductivity but revealed the need of at least $\tau(E,T)$ to reproduce the increased thermopower at low temperatures, $T<\Theta_\text{D}$.

\ack
We want to thank Franziska Maculewicz and Gabi Schierning for enabling us access to detailed low-temperature thermopower measurements of Pb. NFH received funding within the H.C. \O rsted Programme from the European Union's Seventh Framework Programme and Horizon 2020 Research and Innovation Programme under Marie Sklodowska-Curie Actions grant no. 609405 (FP7) and 713683 (H2020).

\bibliographystyle{plain}
\section*{References}
\bibliography{draft_3.bbl}

\begin{thebibliography}{10}

\bibitem{Aynajian:2008}
P.~Aynajian, T.~Keller, L.~Boeri, S.~M. Shapiro, K.~Habicht, and B.~Keimer.
\newblock {Energy Gaps and Kohn Anomalies in Elemental Superconductors}.
\newblock {\em Science}, 319(5869):1509--1512, 2008.

\bibitem{Hellwege:1982:book}
J.~Bass and K.H. Fischer.
\newblock {\em {Metals: Electronic Transport Phenomena}}.
\newblock Landolt-Börnstein, New Series III/15a. 1982.

\bibitem{Bernardi:2014}
Marco Bernardi, Derek Vigil-Fowler, Johannes Lischner, Jeffrey~B Neaton, and
  Steven~G Louie.
\newblock {Ab~Initio Study of Hot Carriers in the First Picosecond after
  Sunlight Absorption in Silicon}.
\newblock {\em Physical Review Letters}, 112(25):257402, June 2014.

\bibitem{Blatt_book:1976}
J~Blatt.
\newblock {\em {Thermoelectric Power of Metals}}.
\newblock Springer-Verlag, 1976.

\bibitem{Bostwick:2007be}
Aaron Bostwick, Taisuke Ohta, Thomas Seyller, Karsten Horn, and Eli Rotenberg.
\newblock {Quasiparticle dynamics in graphene}.
\newblock {\em Nature Physics}, 3(1):36, 2007.

\bibitem{Brown:2016wa}
Ana~M Brown, Ravishankar Sundararaman, Prineha Narang, III William A~Goddard,
  and Harry~A Atwater.
\newblock {Nonradiative Plasmon Decay and Hot Carrier Dynamics: Effects of
  Phonons, Surfaces, and Geometry}.
\newblock {\em ACS Nano}, 10(1):957, 2016.

\bibitem{Cook:1974}
J~G Cook, M~J Laubitz, and M~P Van~der Meer.
\newblock {Thermal conductivity, electrical resistivity, and thermoelectric
  power of Pb from 260 to 550 K}.
\newblock {\em Journal of Applied Physics}, 45(2):510--5, 1974.

\bibitem{DalCorso:2008}
Andrea Dal~Corso.
\newblock {Ab initiophonon dispersions of face centered cubic Pb: effects of
  spin-orbit coupling}.
\newblock {\em Journal of Physics: Condensed Matter}, 20(44):445202--7,
  September 2008.

\bibitem{Fiorentini:2016ki}
Mattia Fiorentini and Nicola Bonini.
\newblock {Thermoelectric coefficients of n-doped silicon from first principles
  via the solution of the Boltzmann transport equation}.
\newblock {\em Phys. Rev. B}, 94(8):085204, 2016.

\bibitem{Floris:2007}
A.~Floris, A.~Sanna, S.~Massidda, and E.~K.~U. Gross.
\newblock {Two-band superconductivity in Pb from ab initio calculations}.
\newblock {\em Phys. Rev. B}, 75:054508, 2007.

\bibitem{Giannozzi:2009}
Paolo Giannozzi, Stefano Baroni, Nicola Bonini, Matteo Calandra, Roberto Car,
  Carlo Cavazzoni, Davide Ceresoli, Guido~L Chiarotti, Matteo Cococcioni,
  Ismaila Dabo, Andrea Dal~Corso, Stefano de~Gironcoli, Stefano Fabris, Guido
  Fratesi, Ralph Gebauer, Uwe Gerstmann, Christos Gougoussis, Anton Kokalj,
  Michele Lazzeri, Layla Martin-Samos, Nicola Marzari, Francesco Mauri,
  Riccardo Mazzarello, Stefano Paolini, Alfredo Pasquarello, Lorenzo Paulatto,
  Carlo Sbraccia, Sandro Scandolo, Gabriele Sclauzero, Ari~P Seitsonen,
  Alexander Smogunov, Paolo Umari, and Renata~M Wentzcovitch.
\newblock {QUANTUM ESPRESSO: a modular and open-source software project for
  quantum simulations of materials}.
\newblock {\em Journal of Physics: Condensed Matter}, 21(39):395502--20,
  September 2009.

\bibitem{Giannozzi:1991}
Paolo Giannozzi, Stefano de~Gironcoli, Pasquale Pavone, and Stefano Baroni.
\newblock {Ab initio calculation of phonon dispersions in semiconductors}.
\newblock {\em Physical Review B}, 43(9):7231--7242, March 1991.

\bibitem{Giustino:2016vf}
Feliciano Giustino.
\newblock {Electron-phonon interactions from first principles}.
\newblock {\em Rev. Mod. Phys.}, 2016.

\bibitem{Gunst:2016}
Tue Gunst, Troels Markussen, Kurt Stokbro, and Mads Brandbyge.
\newblock {First-principles method for electron-phonon coupling and electron
  mobility: Applications to two-dimensional materials}.
\newblock {\em Physical Review B}, 93(3):035414--14, January 2016.

\bibitem{Heid:2010}
R~Heid, I~Yu Bohnen, K P~Sklyadneva, and E~V Chulkov.
\newblock {Effect of spin-orbit coupling on the electron-phonon interaction of
  the superconductors Pb and Tl}.
\newblock {\em Physical Review B}, 81(17):174527--6, May 2010.

\bibitem{Hwang:2008}
E~H Hwang and S~Das~Sarma.
\newblock {Acoustic phonon scattering limited carrier mobility in
  two-dimensional extrinsic graphene}.
\newblock {\em Physical Review B}, 77(11):115449--6, March 2008.

\bibitem{Kaasbjerg:2012uy}
Kristen Kaasbjerg, Kristian Thygesen, and Karsten Jacobsen.
\newblock {Phonon-limited mobility in n-type single-layer MoS$_2$ from first
  principles}.
\newblock {\em Phys. Rev. B}, 85(11):115317, 2012.

\bibitem{Kasinathan:2006}
Deepa Kasinathan, J~Kune{\v s}, A~Lazicki, H~Rosner, C~S Yoo, R~T Scalettar,
  and W~E Pickett.
\newblock {Superconductivity and Lattice Instability in Compressed Lithium from
  Fermi Surface Hot Spots}.
\newblock {\em Physical Review Letters}, 96(4):047004--4, February 2006.

\bibitem{Levy_book:2000}
L.-P. L\'{e}vy.
\newblock {\em {Magnetism and Superconductivity}}.
\newblock Springer-Verlag, 2000.

\bibitem{Li:2015}
Wu~Li.
\newblock { Electrical transport limited by electron-phonon coupling from
  Boltzmann transport equation: An ab initio study of Si, Al, and MoS$_2$}.
\newblock {\em Physical Review B}, 92(7), 2015.

\bibitem{Liao:2015_2}
Bolin Liao, Bo~Qiu, Jiawei Zhou, Samuel Huberman, Keivan Esfarjani, and Gang
  Chen.
\newblock {Significant Reduction of Lattice Thermal Conductivity by the
  Electron-Phonon Interaction in Silicon with High Carrier Concentrations: A
  First-Principles Study}.
\newblock {\em Physical Review Letters}, 114(11):115901--6, March 2015.

\bibitem{Liao:2015ca}
Bolin Liao, Jiawei Zhou, Bo~Qiu, Mildred~S Dresselhaus, and Gang Chen.
\newblock {Ab initio study of electron-phonon interaction in phosphorene}.
\newblock {\em Phys. Rev. B}, 91(23):235419, 2015.

\bibitem{Liao:2015_1}
Bolin Liao, Jiawei Zhou, Bo~Qiu, Mildred~S Dresselhaus, and Gang Chen.
\newblock {Ab initio study of electron-phonon interaction in phosphorene}.
\newblock {\em Physical Review B}, 91(23):235419--8, June 2015.

\bibitem{Liu:1996}
Amy~Y. Liu and Andrew~A. Quong.
\newblock {Linear-response calculation of electron-phonon coupling parameters}.
\newblock {\em Phys. Rev. B}, 53:R7575--R7579, 1996.

\bibitem{Liu:2017jf}
Te-Huan Liu, Jiawei Zhou, Bolin Liao, David~J Singh, and Gang Chen.
\newblock {First-principles mode-by-mode analysis for electron-phonon
  scattering channels and mean free path spectra in GaAs}.
\newblock {\em Phys. Rev. B}, 95(7):075206, February 2017.

\bibitem{Mustafa:2016}
Jamal~I Mustafa, Marco Bernardi, Jeffrey~B Neaton, and Steven~G Louie.
\newblock {Ab initio electronic relaxation times and transport in noble
  metals}.
\newblock {\em Physical Review B}, 94(15):155105, 2016.

\bibitem{Noffsinger:2012va}
Jesse Noffsinger, Emmanouil Kioupakis, Chris~G Van De~Walle, Steven~G Louie,
  and Marvin~L Cohen.
\newblock {Phonon-Assisted Optical Absorption in Silicon from First
  Principles}.
\newblock {\em Phys Rev Lett}, 108(16):7402, 2012.

\bibitem{Park:2014}
Cheol-Hwan Park, Nicola Bonini, Thibault Sohier, Georgy Samsonidze, Boris
  Kozinsky, Matteo Calandra, Francesco Mauri, and Nicola Marzari.
\newblock {Electron Phonon Interactions and the Intrinsic Electrical
  Resistivity of Graphene}.
\newblock {\em Nano Letters}, 14(3):1113--1119, March 2014.

\bibitem{Park:2007kx}
Cheol-Hwan Park, Feliciano Giustino, Marvin~L Cohen, and Steven~G Louie.
\newblock {Velocity Renormalization and Carrier Lifetime in Graphene from the
  Electron-Phonon Interaction}.
\newblock {\em Phys Rev Lett}, 99(8):086804, 2007.

\bibitem{Patrick:2014dp}
Christopher~E Patrick and Feliciano Giustino.
\newblock {Unified theory of electron-phonon renormalization and
  phonon-assisted optical absorption}.
\newblock {\em J. Phys.: Condens. Matter}, 26(36):365503, August 2014.

\bibitem{Piscanec:2004}
S~Piscanec, M~Lazzeri, Francesco Mauri, A~C Ferrari, and J~Robertson.
\newblock {Kohn Anomalies and Electron-Phonon Interactions in Graphite}.
\newblock {\em Physical Review Letters}, 93(18):185503--4, October 2004.

\bibitem{Ponce:2016dv}
S~Ponc{\'e}, E~R Margine, C~Verdi, and F~Giustino.
\newblock {EPW: Electron phonon coupling, transport and superconducting
  properties using maximally localized Wannier functions}.
\newblock {\em Computer Physics Communications}, 209:116--133, December 2016.

\bibitem{Rahim:1975}
C~A Rahim and R~D Barnard.
\newblock {The absolute thermopower of lead above room temperature}.
\newblock {\em Journal of Physics D: Applied Physics}, 8(11):L129--L132, August
  1975.

\bibitem{Roberts}
R~B Roberts.
\newblock {The absolute scale of thermoelectricity}.
\newblock {\em Philosophical Magazine}, 36(1):91--107, August 1977.

\bibitem{Savrasov:1996}
S~Y Savrasov and D~Y Savrasov.
\newblock {Electron-phonon interactions and related physical properties of
  metals from linear-response theory}.
\newblock {\em Physical Review B}, 54(23):16487--16501, 1996.

\bibitem{Schierning}
Gabi Schierning.
\newblock {}.
\newblock {\em private communication}, 2016.

\bibitem{Sklyadneva:2013et}
I~Yu Sklyadneva, R~Heid, K-P Bohnen, P~M Echenique, and E~V Chulkov.
\newblock {Mass enhancement parameter in free-standing ultrathin Pb(111) films:
  The effect of spin-orbit coupling}.
\newblock {\em Phys. Rev. B}, 87(8):085440, February 2013.

\bibitem{Sklyadneva:2012}
I~Yu Sklyadneva, R~Heid, P~M Echenique, K~B Bohnen, and E~V Chulkov.
\newblock {Electron-phonon interaction in bulk Pb: Beyond the Fermi surface}.
\newblock {\em Physical Review B}, 85(15):155115--6, April 2012.

\bibitem{Sofo:1996}
J~O Sofo and G~D MAHAN.
\newblock {\em {The best thermoelectric}}.
\newblock Proc. Natl. Acad. Sci. USA, 1996.

\bibitem{Tandon:2015}
Nandan Tandon, J~D Albrecht, and L~R Ram-Mohan.
\newblock {Electron-phonon interaction and scattering in Si and Ge:
  Implications for phonon engineering}.
\newblock {\em Journal of Applied Physics}, 118(4):045713--6, July 2015.

\bibitem{Xu:2014uh}
Bin Xu and Matthieu Verstraete.
\newblock {First Principles explanation of the positive Seebeck coefficient of
  lithium}.
\newblock {\em Phys Rev Lett}, 112(19):196603, 2014.

\bibitem{Zahn:2011}
Peter Zahn, Nicki~F Hinsche, B~Yu Yavorsky, and Ingrid Mertig.
\newblock Bi2te3: implications of the rhombohedral k-space texture on the
  evaluation of the in-plane/out-of-plane conductivity anisotropy.

\bibitem{Zdetsis:1980}
A~D Zdetsis, E~N Economou, and D~A Papaconstantopoulos.
\newblock {Ab initio bandstructure of lead}.
\newblock {\em Journal of Physics F-Metal Physics}, 10(6):1149--1156, June
  1980.

\bibitem{Ziman_book:1963}
J~M Ziman.
\newblock {\em {Electrons and phonons}}.
\newblock Oxford University Press, 1963.

\end{thebibliography}

\newpage
\pagebreak

\appendix
\section{The electronic nesting function}

\begin{figure}[tbh]
\centering
\includegraphics[width=0.7\columnwidth]{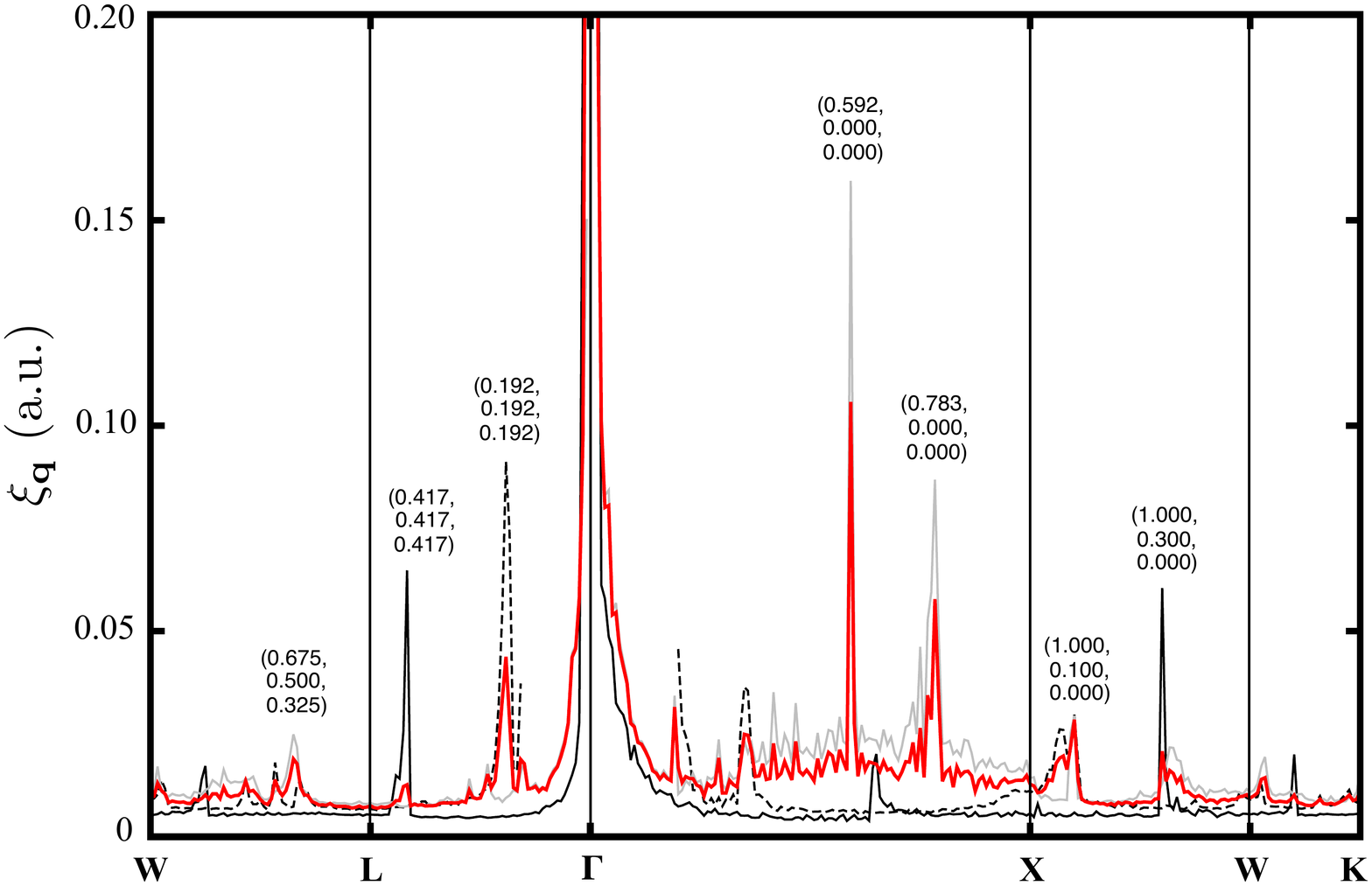}
  \caption{Nesting function $\xi_\mathbf{q}$. The solid black (gray) line corresponds to intraband transitions in the inner (outer) \Fermi sheet, while the dashed black line refers to interband transitions. The total nesting function for the whole \Fermi surface, without distinguishing between intra- and interband transitions, is given in red. In each case the divergence at $\mathbf{q}=\mathbf{0}$ is obvious due to collinear velocities $\mathbf{v}_\kk$ and $\mathbf{v}_{\mathbf{k}+\mathbf{q}}$. Some nesting vectors with larger nesting values are denoted, but the highest values are obtained for vectors not lying on a high-symmetry line, see  figure \ref{fig2_sup}.}
\label{fig1_sup}
\end{figure}

The nesting function \cite{Kasinathan:2006}
\begin{equation}
\xi_\mathbf{q}=\frac{1}{N}\sum_\kk\delkp\delko\propto\oint\frac{d L_\kk}{|\mathbf{v}_\kk\times\mathbf{v}_{\mathbf{k}^\prime}|}
\label{eq:nesting}
\end{equation}
as a part of the \Fermi surfaced averaged \ESF is shown in fig. \ref{fig1_sup}. This quantity indicates phonons from a geometrical point-of-view, which might be important for the correct description of certain properties with respect to the electron-phonon interaction. 
Phonons with large or diverging nesting values connect parallel orientated parts of the \Fermi surface, where the velocities of $\kk$ and $\mathbf{k}^\prime=\kk+\mathbf{q}$ are collinear. This feature is peculiarly pronounced at $\mathbf{q}=\mathbf{0}$, where the velocities of initial and final state are equal. Looking at each \Fermi sheet separately (solid black (grey) line), $\xi_\mathbf{q}$ maximizes at different wave vectors, according to the geometry of sheets, for example $\mathbf{q}^\text{intra}_\text{B1}=(0.417,0.417,0.417)$ or $\mathbf{q}^\text{intra}_\text{B2}=(0.592,0.0,0.0)$. Combining intra- and interband transitions (red line) leads to a drastical reduction of some peaks ($\mathbf{q}^\text{intra}_\text{B1}$) while others remain dominant ($\mathbf{q}^\text{intra}_\text{B2}$). These nesting vectors can be identified via certain cuts through the \Fermi surface and are linked to \textsc{Kohn} anomalies \cite{DalCorso:2008}. 

\begin{figure*}[bt]
\centering
 \begin{minipage}{0.45\textwidth}
 \includegraphics[width=\textwidth]{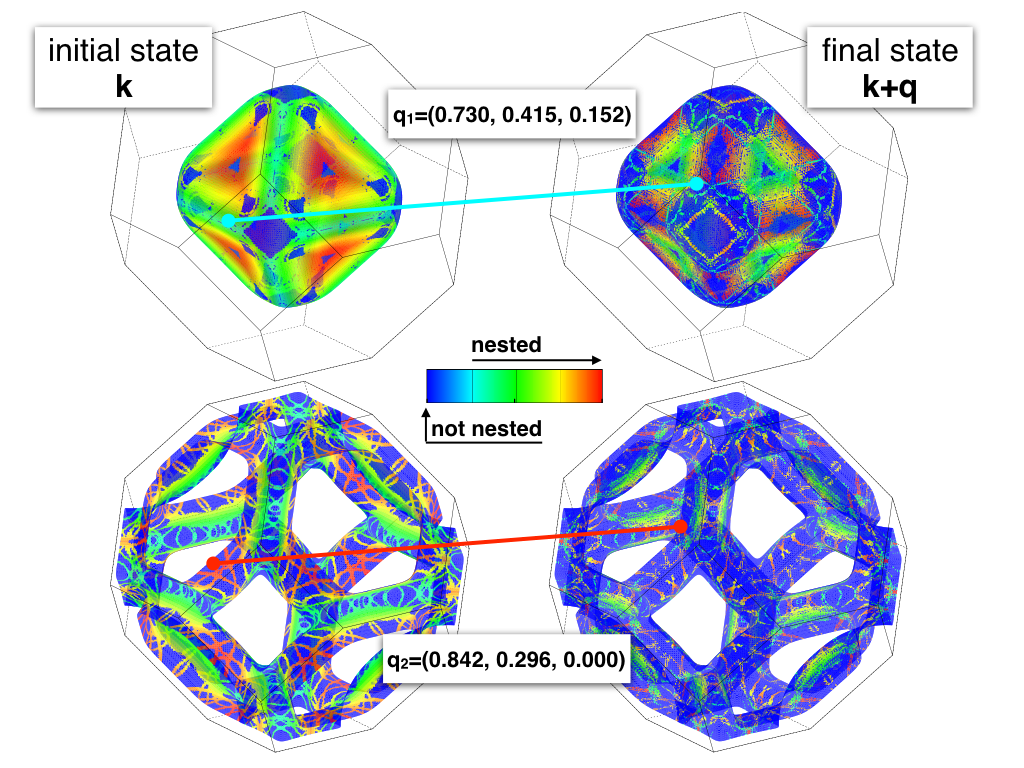}
 \end{minipage}
 \begin{minipage}{0.45\textwidth}
 \includegraphics[width=\textwidth]{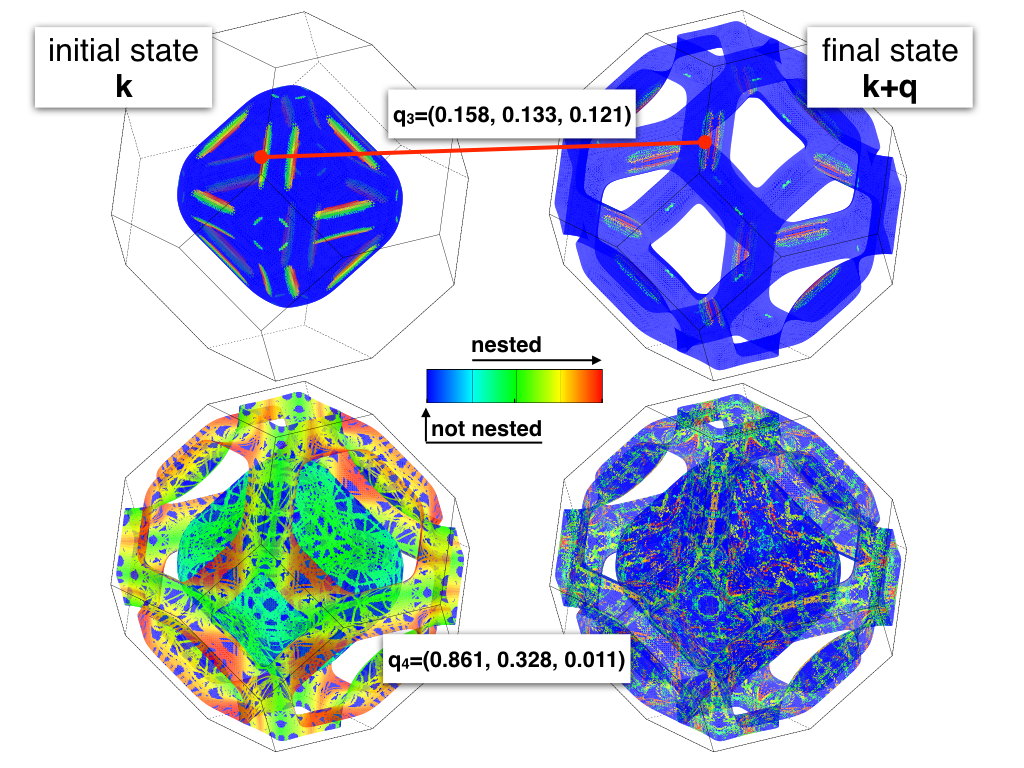}
 \end{minipage}
  \caption{Nested electronic states via certain phonons highlighted at the \Fermi surface. Same color of the initial and final state marks a nested pair, except for blue, which displays the bare \Fermi surface. Only states connected by the phonon with the highest nesting value $\xi_\mathbf{q}$ are shown. If $\mathbf{q}$ connects more than one pair only the pair with the highest weight is given. Left: Intraband transitions mediated by $\mathbf{q}_1$ (top, inner sheet, $\xi_{\mathbf{q}_1}\approx0.495$) and $\mathbf{q}_2$ (bottom, outer sheet, $\xi_{\mathbf{q}_2}\approx0.19$). Right: Interband transitions via $\mathbf{q}_3$ (top, $\xi_{\mathbf{q}_3}\approx0.219$). The bottom figure shows the result, taking both transition types into account ($\xi_{\mathbf{q}_4}\approx0.108$).}
\label{fig2_sup}
\end{figure*}

Besides this more educational investigation of phonons along a high-symmetry line, other wave vectors in the full phonon spectra provide larger nesting values up to a factor of 7 (see fig. \ref{fig2_sup}). This is especially the case for separated intra- and interband transitions. The figures in \ref{fig2_sup} show connected initial and final states via the phonon with the largest nesting (same color). If the phonon couples more than two states only the pair with the highest contribution to $\xi_\mathbf{q}$ is shown. As one can see, the largest nesting vectors lie off the high-symmetry line.
Adding up intra- and interband transitions (bottom right) results in a nesting vector $\mathbf{q}_4$, which provides only a slightly larger nesting than $\mathbf{q}_\text{B2}$. However the visualization of connected states is much more complicated in this case and therefore not shown in the figure.

\newpage

\section{Details of the calculation of $\tau(E,T)$ from $\tau_\kk(T)$}

\begin{figure}[hbt]
\centering
\includegraphics[width=0.7\columnwidth]{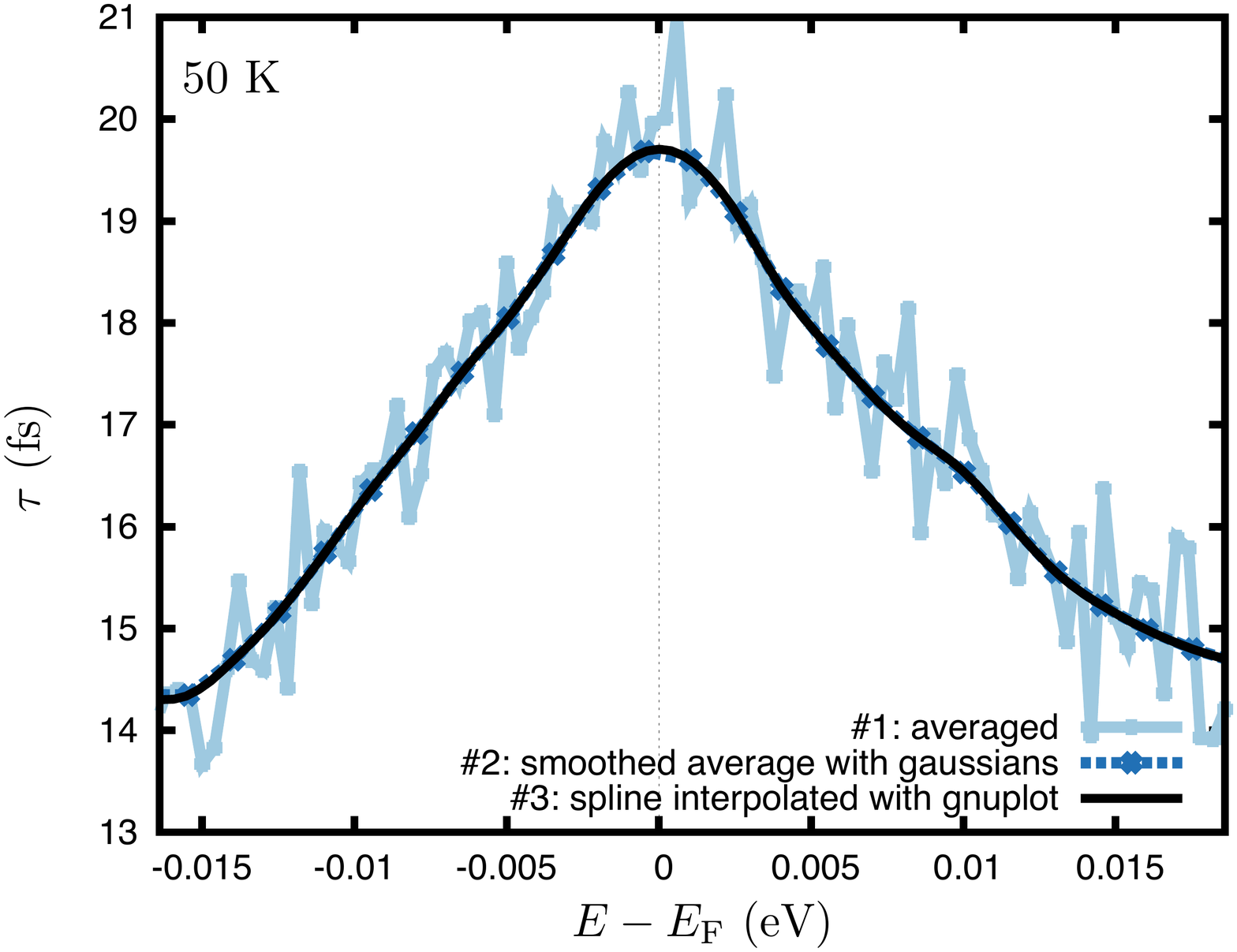}
  \caption{A more detailed picture of the energy-dependent relaxation time at 50K, which is shown in figure 7 in the main text. The energy window is much smaller around the \Fermi energy to show the jagged arithmetic mean value $\tau^\text{AMV}$, which was calculated from state-dependent $\tau_\kk$.}
\label{fig3_sup}
\end{figure}

Figure \ref{fig3_sup} shows the calculation of $\tau(E,T)$ from $\tau_\kk$, which is given in fig. 7 in the main text, in more detail. The arithmetic mean value $\tau^\text{AMV}$ is obviously not smooth but enters directly into the generalized transport coefficients. Because the calculation of the thermopower is really sensitive to the slope of the transport distribution function, which is hardly affected by the relaxation time, $\tau^\text{AMV}$ has to be smoothen before the calculation. After the smoothing with adaptive gaussians, we further use a spline interpolation. Nevertheless, this last step does not change the result significantly. The parameters for the gaussians are listed in table \ref{tab:smear}.

\begin{table}[hbt]
\caption{Used smearing values in the gaussians to smoothen the calculated arithemtic mean values of the relaxation time. Smearing is given in meV.}
\begin{tabular}{cccccc}
$\text{segment}$ & $10 \text{K}$ & $50 \text{K}$ & $100 \text{K}$ & $300 \text{K}$ & $600 \text{K}$\\
\hline
I & 2 & 2 & 20 & 50 & 50 \\
II/III & 50 & 50 & 50 & 50 & 50 \\
\end{tabular}
\label{tab:smear}
\end{table}

\newpage

\section{Comparison of carrier velocity, relaxation time and matrix elements}

\begin{figure}[hbt]
\centering
\includegraphics[width=0.9\columnwidth]{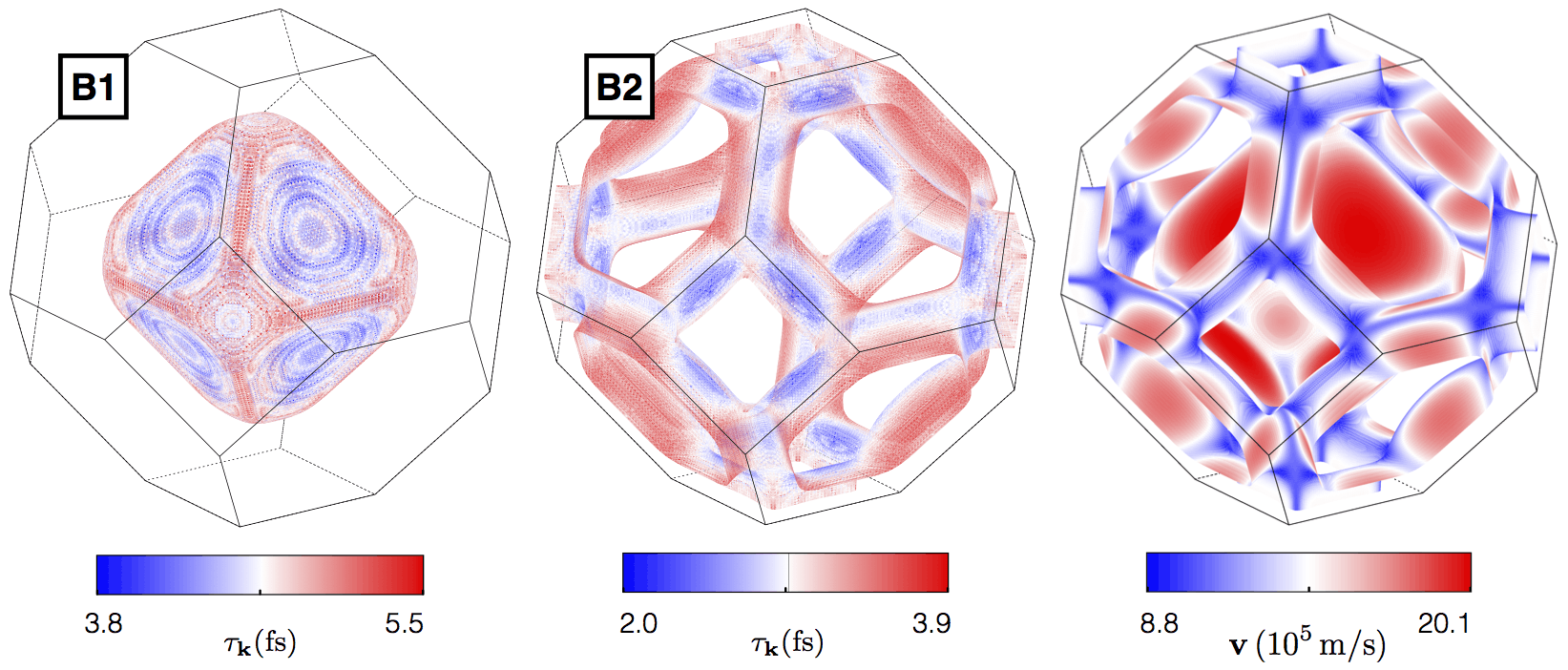}
  \caption{Left and middle: State-dependent relaxation time $\tau_\kk$ for intitial states at the inner \Fermi sheet B1 and the outer \Fermi sheet B2. Take care of the different scales. Right: Carrier velocity in Pb.}
\label{fig4_sup}
\end{figure}

Figure \ref{fig4_sup} indicates, that there is probably no clear correlation between the carrier velocity $\mathbf{v}_\kk$ of the initial state and its relaxation time $\tau_\kk$. If the initial state is at the inner \Fermi sheet (B1), it holds $\tau_\kk\propto\nicefrac{1}{\mathbf{v}_\kk}$, while the opposite is true for an initial state at the outer sheet (B2), where it is $\tau_\kk\propto\mathbf{v}_\kk$.

\begin{figure}[h]
\centering
\includegraphics[width=0.7\columnwidth]{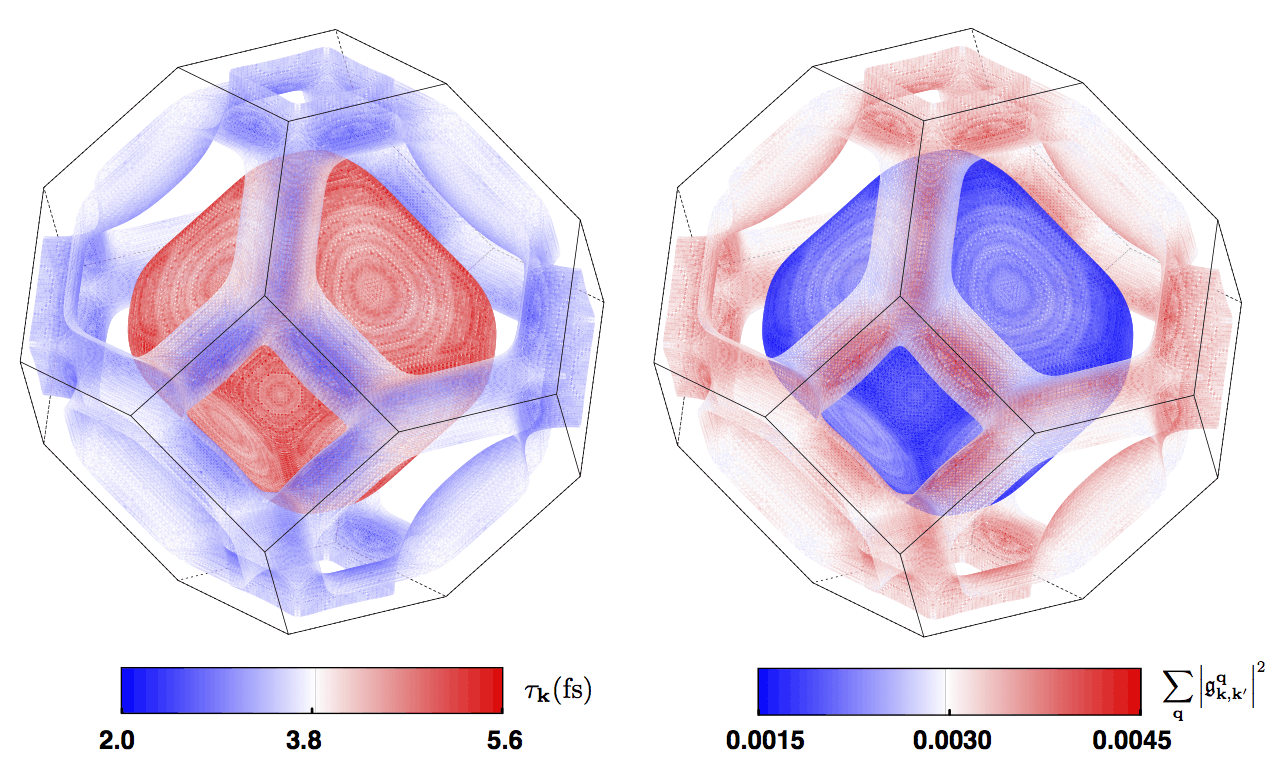}
  \caption{Left: State-dependent relaxation time $\tau_\kk$ for intitial states at the \Fermi surface without distinguishing between intra- and interband scattering. Right: Matrix elements for the same initial states summed over phonons.}
\label{fig5_sup}
\end{figure}

In contrast to the \Fermi velocity, the investigation of the summed up matrix elements $\sum_\mathbf{q}\left|\mathfrak{g}\right|^2$ for each initial state can explain larger or smaller relaxation times, which can be seen in figure \ref{fig5_sup}. Mustafa \textit{et. al} showed this already for simple metals \cite{Mustafa:2016} but it seems to be reasonable in more complicated metals as well.

\end{document}